\documentclass[a4paper,11pt]{article}
\pdfoutput=1 

\usepackage{jheppub} 
\usepackage[T1]{fontenc} 
\usepackage[italian,english]{babel}
\usepackage{hyperref}
\usepackage{ifpdf}
\usepackage{subfigure}
\usepackage{tikz}
\usepackage[compat=1.1.0]{tikz-feynman}
\usepackage{slashed}
\usetikzlibrary{arrows,shapes}
\usetikzlibrary{trees}
\usetikzlibrary{matrix,arrows} 
\usetikzlibrary{positioning}
\usetikzlibrary{calc,through}
\usetikzlibrary{decorations.pathreplacing}  
\usepackage{pgffor}

\usetikzlibrary{decorations.pathmorphing}
\usetikzlibrary{decorations.markings}
\tikzset{
vector/.style={decorate, decoration={snake}, draw},
fermion/.style={draw=black, postaction={decorate}}, 
scalar/.style={dashed,draw=black, postaction={decorate}}}
\tikzstyle{block} = [draw, rectangle, 
minimum height=3em, minimum width=6em]

\usepackage{amssymb}
\usepackage{amsfonts}
\usepackage{epsf}
\usepackage{rotating}
\usepackage{graphicx}
\usepackage{amsmath}
\usepackage{fancyhdr}
\usepackage{lineno}
\usepackage{babel}
\usepackage{graphics}
\usepackage{pstricks}
\usepackage{color}
\usepackage{multirow}
\usepackage{float}
\usepackage{lineno}
\usepackage{mathtools}
\usepackage{stackengine}
\usepackage{comment}

\newcommand{\lsim}{\mathrel{\mathop{\kern 0pt \rlap
{\raise.2ex\hbox{$<$}}}
\lower.9ex\hbox{\kern-.190em $\sim$}}}
\newcommand{\gsim}{\mathrel{\mathop{\kern 0pt \rlap
{\raise.2ex\hbox{$>$}}}
\lower.9ex\hbox{\kern-.190em $\sim$}}}

\newcommand{\be}{\begin{equation}}
\newcommand{\ee}{\end{equation}}
\newcommand{\bea}{\begin{eqnarray}}
\newcommand{\eea}{\end{eqnarray}}

\def\gev{\ensuremath{\mathrm{\,Ge\kern -0.1em V\,}}}
\def\tev{\ensuremath{\mathrm{\,Te\kern -0.1em V\,}}}

\newcommand{\AddrHBNI}{
	Homi Bhabha National Institute, BARC Training School Complex, Anushakti Nagar, Mumbai 400094, India }
\preprint{ CQUeST-2025-0756 \\ \hspace*{0pt}\hfill IP/BBSR/2025-02}

\begin{document} 

\title{ Illuminating Scalar Dark Matter Co-Scattering in EFT with Monophoton Signatures}

\author[a]{Genevi\`{e}ve B\'{e}langer,}
\author[b,c]{Manimala Mitra,}
\author[d]{Rojalin Padhan,}
\author[e,f]{Abhishek Roy}

\affiliation[a]{LAPTh, CNRS, USMB, 9 Chemin de Bellevue, 74940 Annecy, France}
\affiliation[b]{Institute of Physics, Sachivalaya Marg, Bhubaneswar, Odisha 751005, India}
\affiliation[c]{\AddrHBNI}
\affiliation[d]{Department of Physics, Chung-Ang University, Seoul 06974, Korea}
\affiliation[e]{Center for Quantum Spacetime, Sogang University, 35 Baekbeom-ro, Mapo-gu, Seoul, 121-742, South Korea}
\affiliation[f]{Department of Physics, Sogang University, 35 Baekbeom-ro, Mapo-gu, Seoul, 121-742, South Korea}

\emailAdd{belanger@lapth.cnrs.fr}
\emailAdd{manimala@iopb.res.in}
\emailAdd{rojalinpadhan2014@gmail.com}
\emailAdd{abhishek@sogang.ac.kr}


\abstract{We investigate the co-scattering mechanism for   dark matter production in an  EFT framework which contains new $Z_2$-odd singlets, namely 
  two fermions $N_{1,2}$ and a real scalar $\chi$. The singlet scalar $\chi$  is the dark matter candidate.  The dimension-5 operators play a vital role to set the observed DM relic density.
  We focus on a nearly degenerate mass spectrum for the $Z_2$ odd particles to allow for a significant contribution from the   co-scattering or co-annihilation mechanisms. 
   We present two benchmark points where either of the two mechanisms primarily set the DM relic abundance. The main constraint on the model  at the LHC arise from  the ATLAS mono-$\gamma$ search. We obtain the parameter space allowed by the observed relic density and the mono-$\gamma$ search  after performing a scan over the key parameters, the  masses $M_{N_{1,2}}, M_\chi$ and couplings $c_3^\prime, y^\prime_{11,22}$.
   We find the region of parameter space where the relic abundance is set primarily by the co-scattering mechanism while being allowed by the LHC search. We also determine  how the model can be further probed at the HL-LHC via the mono-$\gamma$ signature.

}
\maketitle
\flushbottom

\section{Introduction}
	
Despite strong evidence for dark matter in astrophysics and cosmology, the nature of dark matter (DM) remains unknown~\cite{Bertone:2004pz,Cirelli:2024ssz}.  Moreover searches for new physics  at the high energy or intensity frontiers have not provided clear evidence for physics beyond the standard model. This leaves a large number of possibilities for what dark matter might be. Traditionally the new physics models that were proposed to provide a dark matter candidate and address open  issues with the standard model (SM), with  neutrino mass, or with baryon asymmetry  propose a new weakly interacting particle that is in thermal equilibrium with the standard model bath in the early Universe and   freezes-out as the Universe cools down~\cite{Arcadi:2017kky}. The absence of signals in various astroparticle and collider searches for WIMPs has rekindled the interest in alternative  mechanisms for DM formation and include various modifications of the freeze-out process such as  secluded freeze-out~\cite{Pospelov:2007mp,Feng:2008mu}, DM conversion or co-scattering~\cite{Garny:2017rxs,DAgnolo:2017dbv}, forbidden annihilation~\cite{DAgnolo:2015ujb}, 3-2 annihilation~\cite{Hochberg:2014dra}, assisted freeze-out~\cite{Belanger:2011ww} and interactions between dark sectors~\cite{Belanger:2012vp,Belanger:2021lwd},  decoupled freeze-out~\cite{Chu:2011be}  or processes where thermal equilibrium is not achieved such as in the freeze-in mechanism~\cite{McDonald:2001vt,Hall:2009bx} or with axion dark matter~\cite{Adams:2022pbo}.   These expand the range of possibilities for dark matter masses and interaction strengths as well as for the content of the dark sector thus impacting  the particle  and astroparticle physics signatures that can be expected.  For example when dark matter is feebly interacting and freezes-in collider  signatures are linked to the presence of  long-lived particles~\cite{Feng:2003uy,Belanger:2018sti,Belanger:2022gqc}. 

In  the DM conversion or coscattering mechanism,  DM is weakly interacting but  the WIMP self-annihilation is suppressed  such that the process that drives DM formation relies on the interaction between DM and another almost degenerate particle in the dark sector~\cite{Garny:2017rxs,DAgnolo:2017dbv}. Either coannihilation of a DM particle with another particle of the dark sector~\cite{Binetruy:1983jf,Griest:1990kh} or co-scattering of the DM with the SM into another particle of the dark sector can be the dominant process. These processes typically involve weaker couplings of DM to the SM than in the freeze-out scenario although couplings are assumed to be large enough for DM to be in equilibrium with the SM bath. Several models that rely on this mechanism were studied ~\cite{Cheng:2018vaj,Brummer:2019inq,Junius:2019dci,Garny:2021qsr,Alguero:2022inz,Heeck:2022rep,Heisig:2024mwr,Heisig:2024xbh,DiazSaez:2024dzx,DiazSaez:2024nrq} and some signatures involving long-lived particles were explored~\cite{Alguero:2022inz,DiazSaez:2024dzx}.
In this paper we study the co-scattering mechanism in a simplified extension of the SM which contains  two singlet fermions  and  a new real singlet scalar which we assume to be the dark matter.  We work within an EFT framework and include interactions through dimension 5 operators. Such model was considered in ~\cite{Belanger:2021slj} in the regime where DM is feebly interacting and its  interactions are linked to the generation of neutrino masses.

In this work we explore the regions of parameter space where the singlet and lightest fermion are nearly degenerate in mass such that either co-scattering or coannihilation processes dominate dark matter formation. We assume that the quartic coupling between the scalar singlet and the Higgs is small to evade the strong bounds from direct detection ~\cite{Yu:2024xsy,GAMBIT:2017gge}, hence the quartic interactions do not play a role in DM formation.  Co-scattering is typically dominated by the decay of the singlet fermion to DM and a neutrino.  The DM relic density is obtained after solving two Boltzmann equations for the fermion and the scalar respectively. Because of the small couplings involved, the main signature is not directly linked with DM but results from the production and decay of the heavy fermion into the lightest fermion  and a photon. The signature is challenging because of the low energy of the photon when the two fermions have a small mass splitting. We revisit current LHC constraints arising from the ATLAS monophoton search~\cite{ATLAS:2020uiq} and make projections for the HL-LHC. We show that some parameter space of the model can be probed at the LHC.
	
The structure of this paper is as follows: The model is described in  Section~\ref{sec:model} , Section~\ref{sec:FO} discusses DM formation and describes the computation of the relic density. Section~\ref{sec:collider} presents the analysis relevant for establishing the LHC constraints. Section~\ref{sec:Numerical_Analysis} is devoted to numerical results including  both DM and collider observables. Finally, we conclude with a summary  in Section~\ref{sec:conclusion}.

\section{The Model}\label{sec:model}
We propose an EFT framework where the SM field contents is extended with   two SM singlet  fermions $N_1,N_2$ and a real singlet scalar $\chi$. 
 We impose a $Z_2$ symmetry under which both $N_i$ and $\chi$ are odd, and all the SM fields are even. We consider $\chi$  as the DM candidate.
 
The Lagrangian of this EFT framework up-to d=5 operators  is written as
\begin{eqnarray}
\mathcal{L}_{eff} & = & \mathcal{L}_{SM}+ \frac{c_5}{\Lambda} (\overline{L^c} \tilde{\Phi}) (\tilde{\Phi}^\dagger L)+M_{B} \overline{N^c} N 
+\frac{ y}{\Lambda} \overline{L} \tilde{\Phi} N \chi\\ \nonumber  
&& + \frac{c_1}{\Lambda}  \overline{N^c} N \chi^2 +\frac{c_2}{\Lambda}  \overline{N^c} N \Phi^{\dagger} \Phi    + \frac{c_3}{\Lambda} \overline{N^c} \sigma_{\mu \nu} N B^{\mu \nu} +h.c.\\ \nonumber
&& + \lambda \Phi^{\dagger}\Phi \chi^2+ \beta \chi^4.
\label{eq:genL}
\end{eqnarray} 
 In the above equation $L$, $\Phi$ and $B_{\mu \nu}$ are the SM Lepton doublet, Higgs doublet and hypercharge field strength tensor, respectively; additionally, $\tilde{\Phi}=i\sigma_2 \Phi^\star$ and $N^c=(i\gamma^2\gamma^0) \overline{N}^T$, where $N=(N_1, N_2)^T$. Note that in the above Lagrangian flavor indices are implicit. The couplings $c_5$ is a $3\times3$ matrix, $Y$ is a $3\times2$ matrix and $c_{1,2,3}$ are $2\times2$ matrices in flavor space.
The dipole operator,  $\frac{c_3}{\Lambda} N^T C^{-1} \sigma_{\mu \nu} N B^{\mu \nu}$, plays an important role in the phenomenology, in particular it leads to the decay of $N_2\to N_1 \gamma$ which will give the dominant signature of the model at the LHC. This operator is relevant only when considering at least two singlet fermions, which is why we introduced two singlet fermions. 
Note that the renormalizable Yukawa interaction $\overline{L}\Phi N$ is forbidden due to the $Z_2$ symmetry. Therefore, the field $N_i$ can not participate in the SM neutrino mass generation at tree level via the d=5 Weinberg operator and there is no mixing between SM neutrino and $N_i$. After electroweak symmetry breaking,  $c_2$ contribute to the mass of $N$ as $M_N= M_B+ c_2 v^2/\Lambda$. We redefine the couplings of the d=5 operators as, $c_5^\prime=\frac{c_5}{\Lambda},y^\prime=\frac{y}{\Lambda},c_1^\prime=\frac{c_1}{\Lambda}, c_2^\prime=\frac{c_2}{\Lambda},c_3^\prime=\frac{c_3}{\Lambda}$. For simplicity, we consider all the Wilson co-efficient $c_{i}^{\prime}$ and $y^{\prime}$ to be diagonal except for $c_3^{\prime}$ to be off-diagonal.

We have explored the FIMP dark matter in a similar  model which however featured a different discrete symmetry ~\cite{Belanger:2021slj}. In that model  the additional $\mathcal{Z}_2$ symmetry (that ensures stability of the DM particle $N_3$) forbids the d=5 dipole operator. 

\begin{figure}
	\begin{center}
		
		\includegraphics[height=5cm,width=10cm]{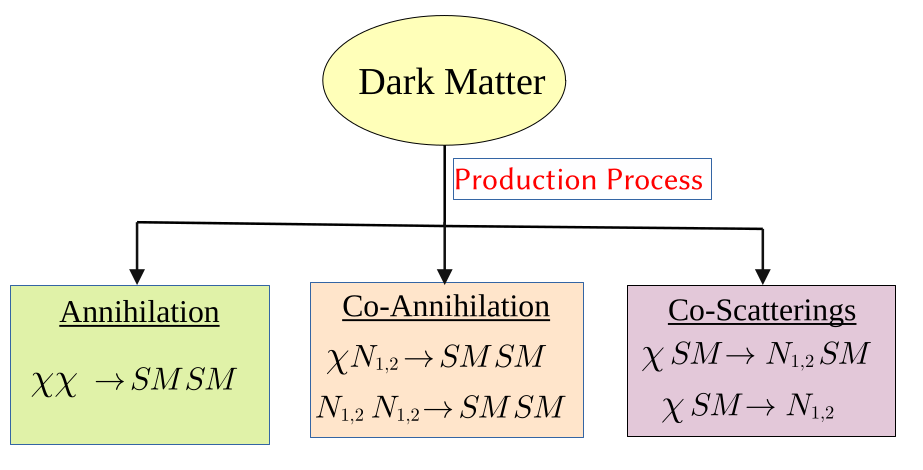}  
		
		\caption{Schematic diagrams representing different WIMP $\chi$ dilution mechanisms in the early Universe for our model.}
		\label{fig:Sch_Diag}
	\end{center}
\end{figure}

\section{Dark Matter Freeze-Out}\label{sec:FO}

We consider both $\chi$ and $N_{1,2}$ to be  in thermal equilibrium with the SM plasma.
The thermal equilibrium of $\chi$ is maintained by the following processes, 
\begin{itemize}
    \item $\chi$ pair annihilation: $\chi \chi \to \text{SM}~\text{SM}$.
    \item $\chi$ and $N_{1,2}$ co-annihilation: $\chi ~N_{1,2} \to \text{SM}~\text{SM}$.
    \item $N_{1,2}/\chi$ interchange: $N_{1,2} ~\text{SM} \to \chi ~ \text{SM}$, $\chi ~ \nu_{e,\mu} \to N_{1,2}$ .
\end{itemize}
 Furthermore, the thermal equilibrium of $N_{1,2}$ is maintained either by the same interactions or by the following processes,
 \begin{itemize}
    \item $N_{1,2}$ pair annihilation: $N_{1,2}~N_{1,2} \to \gamma \gamma,\gamma Z, ZZ$.
    \item $N_{1}$ and $N_{2}$ co-annihilation: $N_{1}~N_{2}\to f\bar{f},W^{+}W^{-},ZH$.
 \end{itemize}
We assume $\chi$ to be the DM particle and $N_{1,2}$ the next-to-lightest-odd (NLOP) particles. 
 We demand $M_{{N}_{1,2}}<1.5M_{\chi}$ to ensure a significant contribution from  either co-annihilation~\cite{Klasen:2013jpa, Molinaro:2014lfa} or co-scattering~\cite{DAgnolo:2017dbv,Junius:2019dci} processes with $\chi$ states. Hence, the abundance of DM $\chi$ is governed by the interplay of $N_{1,2}$ and $\chi$ interactions 
 in the early Universe.
 Schematic diagrams representing different $\chi$ dilution in the early Universe are shown in Fig.~\ref{fig:Sch_Diag}, in addition  $\chi \nu_{e,\mu}\to N_{1,2}$ contribute to co-scattering.

For convenience and to keep with the formalism used in  micrOMEGAs~\cite{Alguero:2022inz} we separate all particles into three sectors, namely, sector $0$ corresponding to SM particles, sector $1$ comprising $N_1$ and $N_2$,  and sector $2$ containing $\chi$~\footnote{Here we will refer to the dark sectors either as sector~1 and  sector~2 or  as $N \,(= N_1 + N_2)$ and $\chi$, respectively.
}. The particles within each sector are assumed to be in thermal equilibrium. The evolution of the dark sector particles in the early Universe is governed by the Boltzmann equations~\cite{Alguero:2022inz},
\begin{eqnarray}
	\frac{dY_1}{dx} &=&-   \frac{1}{x^2}\frac{s(M_\chi)}{\tilde{H}(M_\chi)} \left[    \langle \sigma_{1100} v \rangle ( Y_1^2 - {Y_1^{eq}}^2) +    \langle \sigma_{1122} v \rangle \left( Y_1^2 - Y_2^2  \frac{{Y_1^{eq}}^2}{{Y_2^{eq}}^2}\right) \right. \nonumber\\
	&&+ \left.   \langle \sigma_{1200} v \rangle ( Y_1 Y_2 - Y_1^{eq}Y_2^{eq})
	-\frac{ \Gamma_{2\rightarrow 1}}{s}\left( Y_2 -Y_1 \frac{Y_2^{eq}}{Y_1^{eq}}  \right)        \right] \,,
	\label{eq:Y1}
\end{eqnarray}

\begin{eqnarray}
	\frac{dY_2}{dx} &=&   -\frac{1}{x^2}\frac{s(M_\chi)}{\tilde{H}(M_\chi)}\left[    \langle \sigma_{2200} v \rangle ( Y_2^2 - {Y_2^{eq}}^2) -    \langle \sigma_{1122} v \rangle \left( Y_1^2 - Y_2^2  \frac{{Y_1^{eq}}^2}{{Y_2^{eq}}^2}\right) \right. \nonumber \\
	&&+ \left.  \langle \sigma_{1200} v \rangle ( Y_1 Y_2 - Y_1^{eq}Y_2^{eq})   + \frac{ \Gamma_{2\rightarrow 1}}{s}\left( Y_2 -Y_1 \frac{Y_2^{eq}}{Y_1^{eq}}  \right)        \right] \,.
	\label{eq:Y2}
\end{eqnarray}
Here $Y_{1,2}$ is the abundance of the particles in the  dark sector 1 and 2, and $Y_{1,2}^{eq}$ is the  corresponding equilibrium abundance. $\langle\sigma_{ijkl} v\rangle$ indicates the thermal average cross-section for the process $x_i x_j\to x_k x_l$ where $x_i$ stands for any particle in sector i.  The term $\Gamma_{2\to1}$ is the conversion term, including both co-scattering as well as decay terms, and is given by,
\begin{eqnarray}\label{eqn:coscat}
    \Gamma_{2\rightarrow 1}  =
    \frac{ \sum_{\alpha=1}^{2} \Gamma_{N_{\alpha}\to \chi,\nu_{e,\mu}}\,  g_{\alpha} M_{N_\alpha}^2 K_1\left(\frac{M_{N_\alpha}}{M_{\chi}}x\right)}
           { \sum_{\alpha=1}^{2} g_{\alpha} M_{N_\alpha}^2 K_2\left(\frac{M_{N_\alpha}}{M_{\chi}}x\right) }
    +  \langle \sigma_{2010} v \rangle s(M_\chi/x)Y^{eq}_0 \,,
\end{eqnarray}
where $x=M_\chi/T$, $g_\alpha(=2)$ is the internal degree of freedom of $N_{1,2}$, $K_1,K_2$ are   the modified Bessel function of the second kind of order one and two and $Y^{eq}_0(=0.238)$ is the SM sector comoving equilibrium number density. Note that the modified Hubble rate and entropy density in terms of $M_\chi$ is given by,
\begin{eqnarray}
\tilde{H}(M_\chi)=H(M_\chi)\left[1-\frac{x}{3g^{s}_{*}} \frac{dg^{s}_{*}}{dx}\right]^{-1},\,\,\,\,\,\, s(M_\chi)=\frac{2\pi^2}{45}g^{s}_{*}M_\chi^3,
\label{eq:Hrate}
\end{eqnarray}
where $g^{s}_{*}$ is the effective number of degrees of freedom related to the entropy density of the Universe, $s$.

We used micrOMEGAs~\cite{Alguero:2022inz,Alguero:2023zol} to solve Eq.~\ref{eq:Y1} and Eq.~\ref{eq:Y2}. For this  we implemented the model in Feynrules~\cite{Alloul:2013bka} and generated the model files for CalcHEP~\cite{Belyaev:2012qa} required by micrOMEGAs. The code then solves Eq.~\ref{eq:Y1} and \ref{eq:Y2} for the abundances and  the relic density  given by,
\begin{eqnarray}
\Omega_{\chi} h^{2}=2.742\times10^{8}(M_{N_1}Y_{1}+M_{\chi}Y_{2}).
\label{eq:obs_relic}
\end{eqnarray}
In our analysis, we assume $M_{N_2}>M_{N_1}$, and thus $N_2$ decouples prior to $N_1$ in sector 1. Consequently, at the late epoch of the Universe, only $N_1$ contributes to $Y_1$.  The NLOP $N_1$   decays to DM $\chi$  after $N_1$ freeze-out, i.e $Y_{1}\to 0$ at $T<<M_{N_{1,2}}$, such that the only relevant term that contributes to the observed relic density is from the second term $Y_2$ in Eq.~\ref{eq:obs_relic}. 

The DM relic density is precisely determined by the PLANCK satellite through measurements of the cosmic microwave background (CMB)~\cite{Planck:2018vyg},
\begin{eqnarray}
        \Omega_{\chi}h^{2}=0.12 \pm 0.0012.
\end{eqnarray}
In the numerical analysis we will require that the DM relic density lies  within $2\sigma$ of the observed value.

\begin{figure}
	\begin{center}
		{\includegraphics[width=0.8\linewidth]{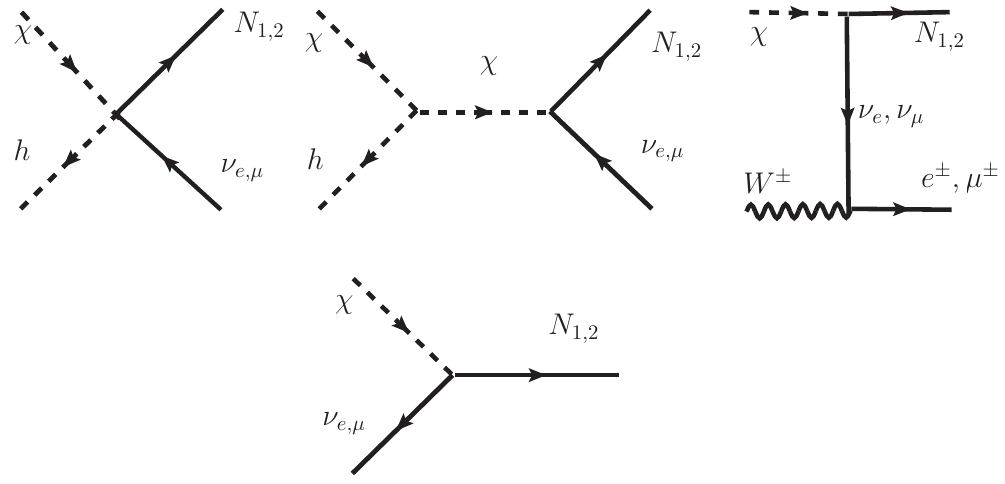}} 
		\caption{Feynman diagrams for the conversion processes that depend on the Yukawa coupling $Y^\prime$. The inverse decay shown in the bottom line usually dominates. }
		\label{fig:S1_diag2}
	\end{center}
\end{figure}
\begin{table}
	\centering
	\renewcommand\arraystretch{1.5}
	\begin{tabular}{| c | c | c | c | c | }
		\hline 
		\multicolumn{2}{|c|}{Initial state} & \multicolumn{2}{|c|}{Final state} & Scaling with couplings \\
		\hline
		\multirow{1}{*}{$\chi$} & \multirow{1}{*}{$h$} & $N_{1,2}$ & $\nu_{e,\mu}$ & \multirow{1}{*}{${y^{\prime}}_{11(22)}^2$} \\ 
		\hline
		\multirow{1}{*}{$\chi$} & \multirow{1}{*}{$W^{\pm}$} & $N_{1,2}$ & $e^{\pm}\mu^{\pm}$ & \multirow{1}{*}{${y^{\prime}}_{11(22)}^2$\,(\textbf{t-channel process }) } \\
		\hline
		\multirow{1}{*}{$\chi$} & \multirow{1}{*}{$\nu_{e,\mu}$} & $N_{1,2}$ & $-$ & \multirow{1}{*}{${y^{\prime}}_{11(22)}^2$\,(\textbf{Inverse Decay}) } \\
		\hline
	\end{tabular}
	\caption{List of all relevant conversion processes shown in Fig.~\ref{fig:S1_diag2}. All  cross sections  depend on the square of the Yukawa coupling $y^\prime$.}
	\label{tab:Conversion_Process}
\end{table}
 
  To illustrate the importance of the co-scattering and co-annihilation mechanisms for obtaining the correct relic density, we introduce $\Delta^{1}_{\chi}$, which is the fractional difference between  the relic densities obtained after solving the  coupled and single Boltzmann equation~\cite{Alguero:2022inz}:
\begin{equation}    
\Delta^{1}_{\chi} \equiv 1-\frac{ \Omega h^2(\textrm{Single}) }{ \Omega h^2(\textrm{Coupled}) }.
\label{eq:fsectors}
\end{equation}
More specifically, $\Omega h^2(\text{Single})$ refers to the value obtained using the standard {\bf darkOmega} function in micrOMEGAs, which accounts for only one Boltzmann equation and assumes that all particles are in thermal equilibrium in the early Universe. Thus it incorporates solely co-annihilation processes. In contrast, $\Omega h^2(\text{Coupled})$ represents the result obtained via the {\bf darkOmegaN} function, where the dark particles are divided into two sectors  that are not  necessarily in thermal equilibrium with each other. This calculation involves solving the coupled system of Boltzmann equations~\ref{eq:Y1} and \ref{eq:Y2}, including all co-scattering and decay processes.

\begin{figure}
	\begin{center}
		{\includegraphics[width=0.8\linewidth]{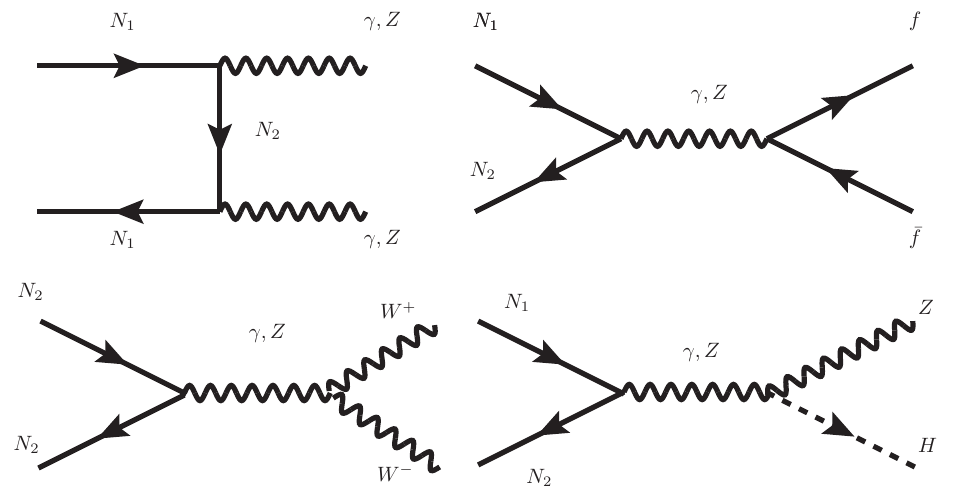}} 
		\caption{Feynman diagrams for the pair annihilation and co-annihilation process of $N_{1,2}$ that depend on the coupling $c_3^\prime$.  Note that due to the choice $(c^{\prime}_{3})_{ij}=\epsilon_{ij}(c^{\prime}_{3})_{ij}$, we have both co-annihilation and annihilation  processes between $N_{1}$ and $N_{2}$.}
		\label{fig:S1_diag1}
	\end{center}
\end{figure}
\begin{table}[]
	\centering
	\renewcommand\arraystretch{1.5}
	\begin{tabular}{| c | c | c | c | c | }
		\hline 
		\multicolumn{2}{|c|}{Initial state} & \multicolumn{2}{|c|}{Final state} & Scaling with couplings \\
		\hline
		\multirow{1}{*}{$N_{1,2}$} & \multirow{1}{*}{$N_{1,2}$} & $\gamma,Z$ & $\gamma,Z$ & \multirow{1}{*}{${c_{3}^{\prime}}^4$\,(\textbf{t- channel process }) } \\
		\hline
		\multirow{1}{*}{$N_{1}$} & \multirow{1}{*}{$N_{2}$} & $f$ & $\bar{f}$ & \multirow{1}{*}{${c_{3}^{\prime}}^2$\,(\textbf{s-channel process }) }  \\
		\hline
        \multirow{1}{*}{$N_{1}$} & \multirow{1}{*}{$N_{2}$} & $W^+$ & $W^-$ & \multirow{1}{*}{${c_{3}^{\prime}}^2$\,(\textbf{s-channel process }) }  \\
		\hline
        \multirow{1}{*}{$N_{1}$} & \multirow{1}{*}{$N_{2}$} & $Z$ & $H$ & \multirow{1}{*}{${c_{3}^{\prime}}^2$\,(\textbf{s-channel process }) }  \\
		\hline
	\end{tabular}
	\caption{List of all relevant annihilation and co-annihilation processes as in Fig.~\ref{fig:S1_diag1}, with the  dependency of their cross sections on the $c_{3}^\prime=(c_{3}^{\prime})_{12}$.  }
	\label{tab:N_1_N_2_c3_dependences}
\end{table}

Note that both sectors remain in thermal equilibrium when the Wilson coefficient ($c_{i}^{\prime},\, y^{\prime}$) and the Higgs portal coupling $\lambda$ are sufficiently large. If this is the case  $Y_{1}/Y_{2}=Y_1^{eq}/Y_2^{eq}$ and Eq.~\ref{eq:Y1},~\ref{eq:Y2} simplify considerably and solving the coupled Boltzmann equations yields the same result as solving a single abundance equation where $Y_1= 0$ and $Y_2= Y$. Hence, the usual freeze-out result is recovered and  $\Delta^{1}_{\chi}=0$. Conversely, in  scenarios where   $y^{\prime}, \lambda \sim \mathcal{O}(10^{-6}-10^{-10})$, $\chi$'s pair annihilation, co-annihilation, and exchange process involving $\chi$ and constituents of sector 2 become negligible, thereby simplifying the Boltzmann equations\footnote{For the numerical analysis, we solve the general Boltzmann equations governing the evolution of $\chi$ and the constituents of sector 1, specifically Eqs.~\ref{eq:Y1} and \ref{eq:Y2}.}, 
\begin{eqnarray}
	\frac{dY_1}{dx} =-   \frac{1}{x^2}\frac{s(M_\chi)}{\tilde{H}(M_\chi)} \left[    \langle \sigma_{1100} v \rangle ( Y_1^2 - {Y_1^{eq}}^2)
	-\frac{ \Gamma_{2\rightarrow 1}}{s}\left( Y_2 -Y_1 \frac{Y_2^{eq}}{Y_1^{eq}}  \right)        \right] \,,
	\label{Ueq:Y1}
\end{eqnarray}

\begin{eqnarray}
	\frac{dY_2}{dx} = -\frac{1}{x^2}\frac{s(M_\chi)}{\tilde{H}(M_\chi)}\left[     \frac{ \Gamma_{2\rightarrow 1}}{s}\left( Y_2 -Y_1 \frac{Y_2^{eq}}{Y_1^{eq}}  \right)        \right] \,.
	\label{Ueq:Y2}
\end{eqnarray}

The dominant  process in the above equation contributing to $\Gamma_{2\to1}$  arises from $N_{1,2}$ decay while $2\to2$ processes are  subdominant. The corresponding Feynman diagrams are presented in Fig.~\ref{fig:S1_diag2}. All cross-sections depend on $y^{\prime 2}$ as shown in Table~\ref{tab:Conversion_Process}. It is important to emphasize that we need at least $N_1$ to be in chemical equilibrium with the thermal bath through (co-)annihilation process, i.e. 1100, which depends on $c_{3}^{\prime}$ solely, to facilitate $\chi$ dilution either through co-annihilation or co-scattering. Note that 1100 refers to processes of the type $N_i N_j\to SM,SM$, these are illustrated in Fig.~\ref{fig:S1_diag1}. Therefore, it becomes necessary to assume $c_{3}^{\prime} >10^{-5}$ to ensure that the constituents of sector 1 remain in chemical equilibrium at $T\geq M_{N_{1,2}}$. On the contrary, due to the small coupling of $\chi$ with $N_{1,2}$ in this scenario, $\chi$ may undergo chemical decoupling at much larger $T$. The relic density of $\chi$  is set when the conversion or co-scattering processes shut off and it freezes out. Furthermore, as we focus on the conversion/co-annihilation processes,  it is convenient to define the following variables, which dictate the relative mass splitting between the dark sector particles,
\begin{eqnarray}
	\delta_1=\frac{{M_{N_1}}-M_\chi}{M_\chi},\,\,\,\,\,\,\delta_2=\frac{{M_{N_2}}-{M_{N}}_{1}}{{M_{N_1}}}
	\label{eq:relative_mass_splitting}
\end{eqnarray}

Before focusing on the primary analysis of this work, we want to bring to the readers' attention  that  $\chi$ pair annihilation through the Higgs portal coupling $\lambda$ is stringently constrained from direct detection (DD) experiments such as 
LUX-Zeplin~\cite {Lz:2024zvo}. Detailed analysis of  $\chi$ pair annihilation analysis through the Higgs portal coupling $\lambda$ can be found in ~\cite{Escudero:2016gzx,GAMBIT:2017gge,Yu:2024xsy,Belanger:2024wca,Escudero:2025eid}. To evade the stringent constraint on $\lambda$ from DD searches while ensuring that  $\chi$ remains in thermal equilibrium, we fix $\lambda$ to $10^{-6}$ such that,
\begin{eqnarray}
\left(\frac{\Gamma_{\chi\chi\to h\to \text{SM}\text{SM}}}{H}\right)_{T=M_{\chi}} \simeq \mathcal{O}(1-100),
\end{eqnarray} 
where $\Gamma_{\chi\chi\to h\to \text{SM}\text{SM}}$ is the $\chi$ pair annihilation rate in the early Universe through the Higgs portal coupling $\lambda$.

In this work, we focus primarily on the thermal production of $\chi$ in a scenario where $c_3^{\prime}$ and $y^{\prime}$ are non-zero, while fixing $\lambda = 10^{-6}$ and setting all other Wilson coefficients to zero\footnote{Assuming $c_{1,2} \neq 0$, $y' \neq 0$, and $c_3^\prime = 0$, similar DM phenomenology may arise under the current assumptions of this work. A thorough exploration of additional scenarios is deferred to future work.}. 
In this case, the value of $\delta_{2}$ determines the dominant process contributing to $N_1$ annihilation. When $\delta_{2} <0.5$, $N_1$ is primarily diluted through co-annihilation while pair annihilation is subdominant; otherwise, pair annihilation of $N_1$ is the dominant process~\cite{Masso:2009mu,Weiner:2012cb,Tulin:2012uq,Cline:2012bz}. 
Note that the dominant co-annihilation process is $N_{1}N_{2}\to f\bar{f}$. In comparison, the process  $N_{1}N_{2}\to W^{+}W^{-}$ contributes less than 1\% to the total co-annihilation rate due to a cancellation between the virtual $Z$ and $\gamma$ exchange diagrams. The relevant Feynman diagrams for $N_{1}$ dilution in the early Universe are shown in Fig.~\ref{fig:S1_diag1}, and the interaction rates dependence on $c_3^{\prime}$ are given in Table~\ref{tab:N_1_N_2_c3_dependences}. It is important to bring to the attention of the readers that a wide range of DM masses is feasible in our model, spanning from sub-GeV to the TeV scale. In our work, we primarily focus on weak-scale DM mass to have a viable collider signature in the LHC, as will be discussed in the next section. In contrast, neither upcoming direct nor indirect detection experiments can provide viable signatures, since the coupling responsible for dark matter detection, $\lambda$, is set to $10^{-6}$.

We begin by examining two benchmark points to illustrate how the evolution of DM $\chi$ depends on the key parameters, i.e. $c_{3}^{\prime},\, y^{\prime}, M_{\chi}\, \text{and}\, M_{N_{1,2}}$.   A detailed analysis of the allowed parameter space will be presented in Section~\ref{sec:Numerical_Analysis}. Based on the primary production mechanism of $\chi$, either through co-annihilation or co-scattering, we have considered two benchmark points. We assume $c_{3}^{\prime}=7\times 10^{-4}~\text{GeV}^{-1}$ and the other parameters are set to, 
\begin{itemize}
    \item {\bf{BP1:}} $M_{\chi}=100\,\text{GeV},\,M_{N_{1}}=105\,\text{GeV},\,M_{N_{2}}=115\,\text{GeV},\,{y^{\prime}}_{11}=y^{\prime}_{22}=1.1\times10^{-9}~\text{GeV}^{-1}$
    \item {\bf{BP2:}} $M_{\chi}=300\,\text{GeV},\,M_{N_{1}}=315\,\text{GeV},\,M_{N_{2}}=347\,\text{GeV},\,{y^{\prime}}_{11}=y^{\prime}_{22}=1.2\times10^{-9}~\text{GeV}^{-1}$
\end{itemize}

\begin{figure}
	\begin{center}
		\mbox{
			\subfigure[]{\includegraphics[width=0.5\linewidth]{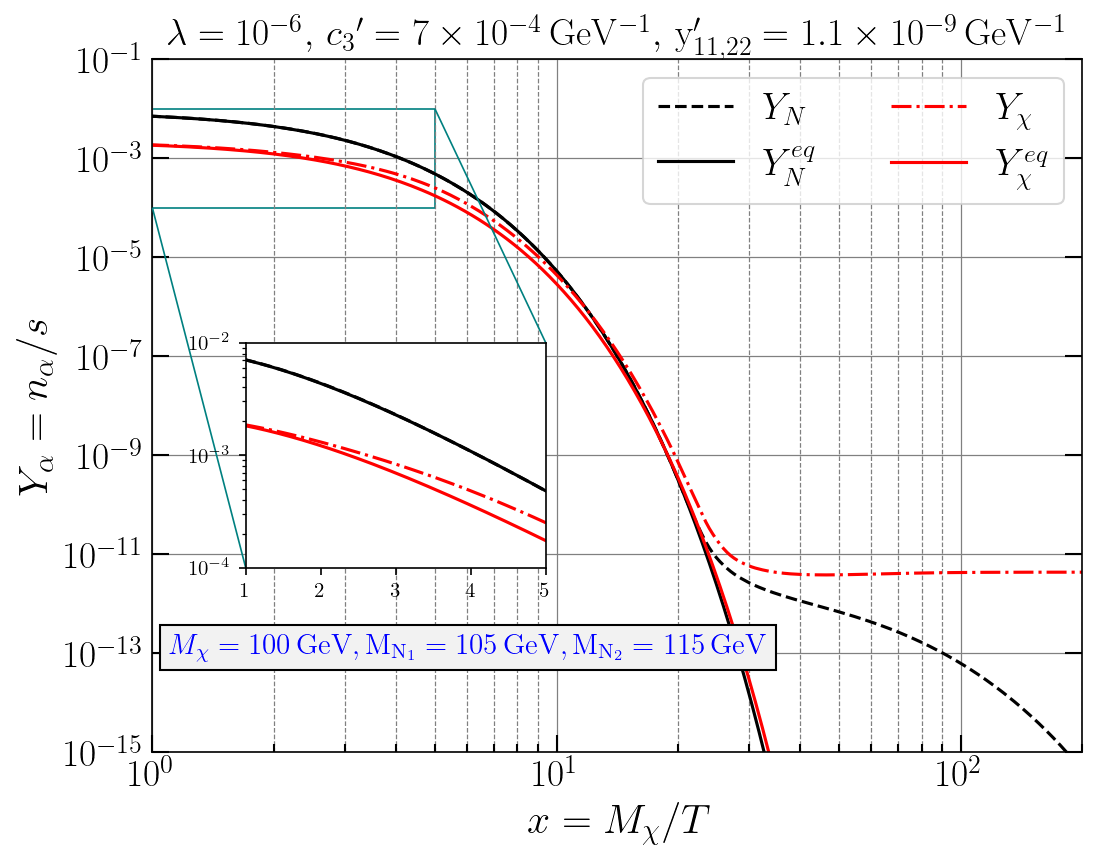}
				\label{fig:S1_DM_yield_1a}}  \subfigure[]{\includegraphics[width=0.5\linewidth]{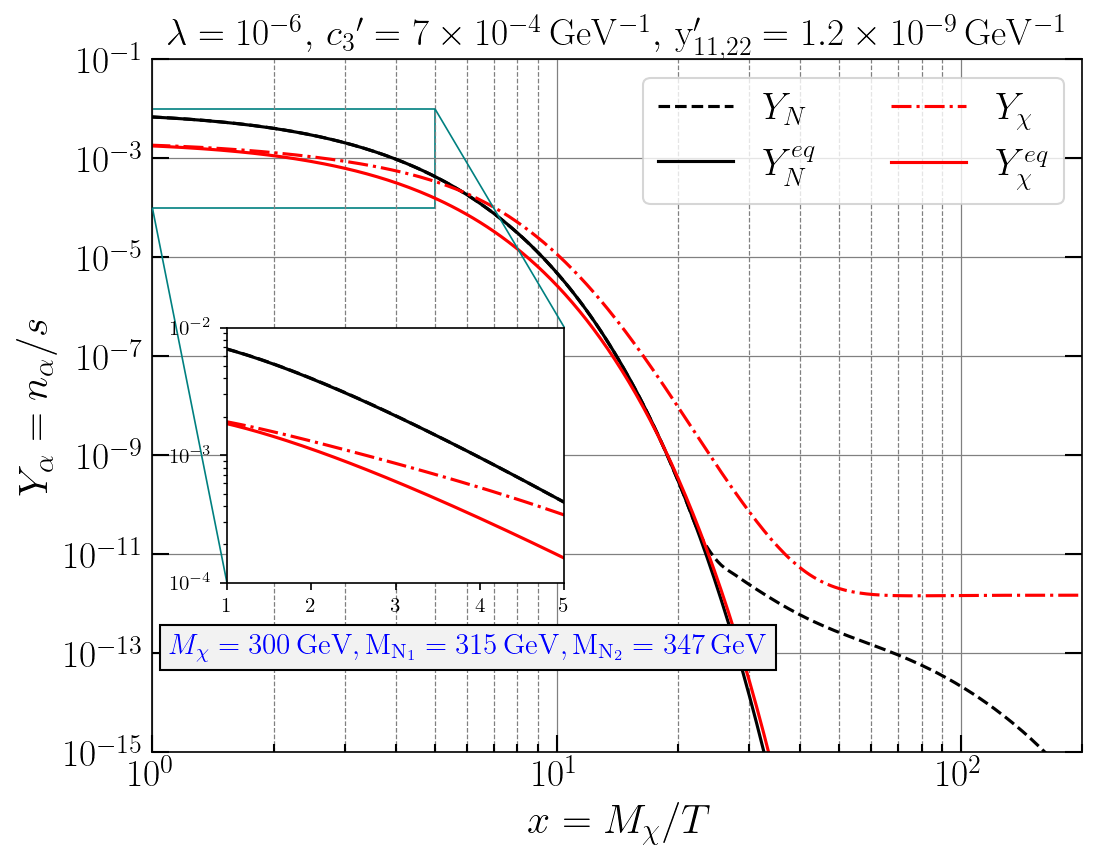}
				\label{fig:S1_DM_yield_1b}}
}
		\caption{Evolution of the $\chi$ and $N(=N_{1}+N_{2})$ abundances in the early universe for the benchmark point (a) BP1 and (b) BP2. The observed DM relic density ($\Omega_{\chi}h^{2} \simeq0.12$) is satisfied for both benchmark points.}
		\label{fig:S1_DM_yield}
	\end{center}
\end{figure}

\begin{figure}
	\begin{center}
		\mbox{
			\subfigure[]{\includegraphics[width=0.5\linewidth]{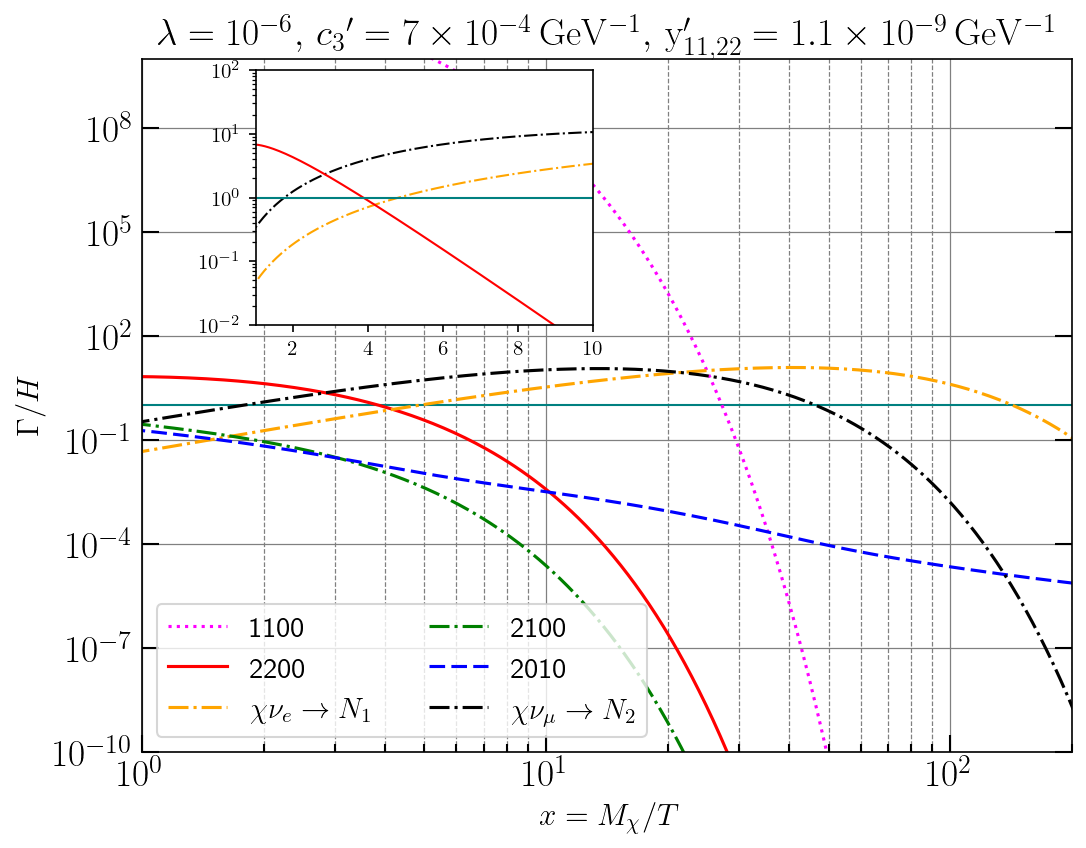}\label{fig:S1_DM_Rate_1a}}  
			\subfigure[]{\includegraphics[width=0.5\linewidth]{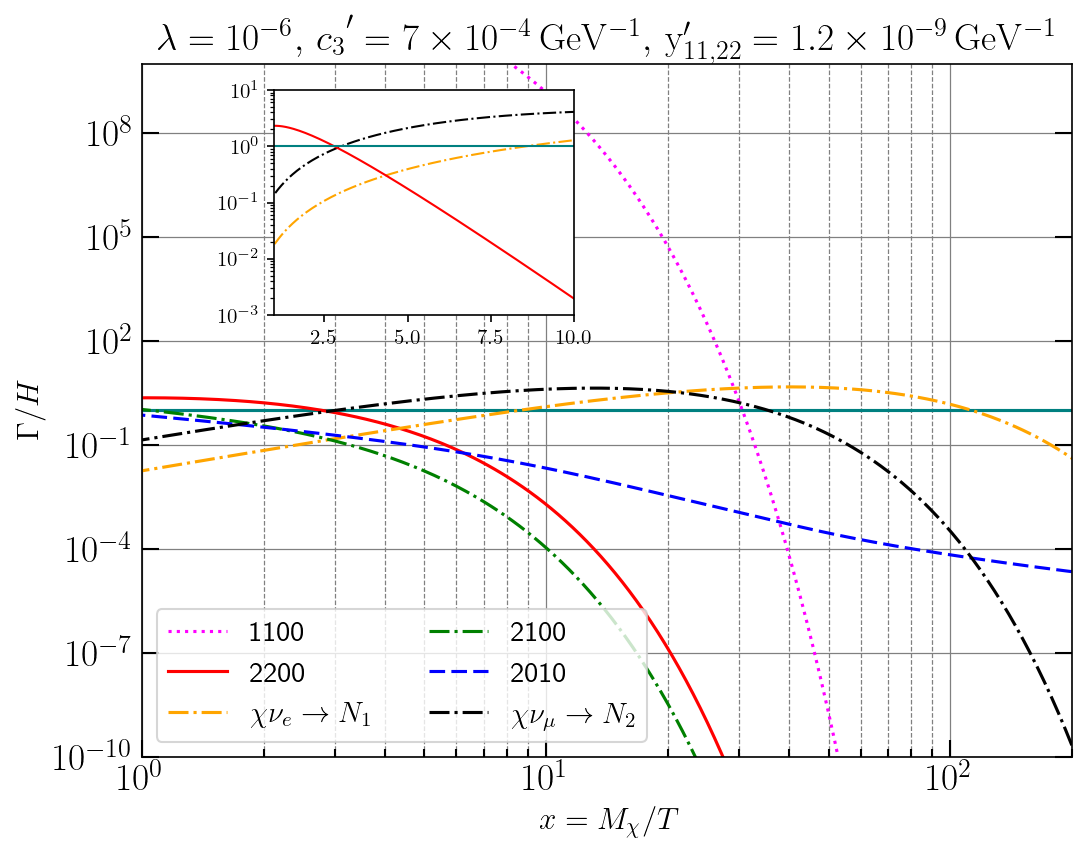}\label{fig:S1_DM_Rate_1b}}		 }
		\caption{The ratio of interaction rates  to the Hubble rate $\Gamma/H$ for various processes  as a function of $x=M_{\chi}/T$ for the benchmark point (a) BP1 and (b) BP2.  The  horizontal light blue  line depicts $\Gamma/H$=1.}
		\label{fig:S1_DM_Rate}
	\end{center}
\end{figure}

Fig.~\ref{fig:S1_DM_yield} shows the evolution of the densities of $\chi$  for  the two benchmark points. Fig.~\ref{fig:S1_DM_yield_1a} and ~\ref{fig:S1_DM_yield_1b}  illustrate that the DM can chemically decouple at early times ($2 \lesssim x \lesssim4 $) and still freeze-out at the late epoch of the Universe. On the contrary, $N_{1,2}$ stays in  chemical equilibrium with the thermal bath until the freeze-out occurs. Note that $N_{1}$ rapidly decays to the DM after freeze-out. However, post freeze-out of $N_1$, since $Y_{N_{1}}<Y_{\chi}$, the decay term gives only a negligible contribution to the DM relic density. 

The departure of $Y_{\chi}$ from chemical equilibrium at an early epoch, as shown in Fig.~\ref{fig:S1_DM_yield_1a} and Fig.~\ref{fig:S1_DM_yield_1b}, arises due to $\chi$ pair annihilation, co-annihilation and $2\to2$ inelastic processes going out of chemical equilibrium at the early epoch of the Universe. To illustrate this in more detail, we show in Fig~\ref{fig:S1_DM_Rate} how $\Gamma/H$ varies as a function of $x$ for different processes for both benchmark points. We can see that all the $2\to2$ processes involving $\chi$ decouple at high temperature (around $x \simeq 4$ for BP1 and $x \simeq 2.5$ for BP2 ). Consequently, $\chi$ undergoes early chemical decoupling in both BP1 and BP2 and ceases to track its equilibrium number density. Moreover, the inverse decay processes $\chi \nu_{e,\mu} \to N_{1,2}$ rate satisfies the following condition~\cite{Frumkin:2021zng,Frumkin:2025dxq},
\begin{eqnarray}
\Gamma_{N_{1,2}\to\chi \nu_{e,\mu}} \lesssim \left(\frac{M_{N_{1,2}}-M_{\chi}}{M_{\chi}} \right)H(x=1),
\label{eqn:inv_decay_condition}
\end{eqnarray}
where $\Gamma_{N_{1,2}\to\chi \nu_{e,\mu}}$ is the partial decay width of $N_{1,2}$ to $\chi$. It ensures that inverse decay processes are too inefficient to restore chemical equilibrium, resulting in $Y_{\chi} > Y^{eq}_{\chi}$, note that $Y_\chi$ remains  close to the equilibrium abundance. Furthermore, for BP1, $\chi \nu_{e,\mu} \to N_{1,2}$ both contribute significantly for $x\gtrsim 2$ which can be seen from the inset of Fig~\ref{fig:S1_DM_Rate_1a}  thus ensuring DM freeze-out is primarily governed by $N_{1}$ freeze-out rather than by inverse decay freeze-out. In contrast, for BP2, the DM decouples when the inverse decay $\chi \nu_{e} \to N_{1}$ ceases.

It is important to point out  that we have streamlined the calculation  by assuming that $\chi$ remains in kinetic equilibrium, thus  we solved the fully integrated Boltzmann equations.  
In our analysis, the inverse decay process dominates over the  $2 \leftrightarrow 2$ interactions. Consequently, the results obtained 
using the fully integrated Boltzmann equations are expected to differ 
by less than $10\%$ from those of the coupled un-integrated Boltzmann 
equations~\cite{Garny:2017rxs,Alguero:2022inz}. This small deviation arises because the decay and inverse decay processes 
remain efficient until late times, thereby re-establishing an approximate kinetic equilibrium (see Fig.~8 of Ref.~\cite{Garny:2017rxs}).
However, in scenarios where the inverse decay becomes subdominant, the deviation  from the coupled un-integrated Boltzmann results can become significant,  as the dark matter momentum distribution becomes highly distorted (i.e., deviates considerably from a thermal distribution). In such cases,  the integrated Boltzmann equation may under- or overestimate the actual relic yield~\cite{Brummer:2019inq}, thus solving  the  coupled un-integrated
Boltzmann equation will provide a more accurate estimate of the relic abundance of
the DM. We leave this for future work.
\section{Collider Constraints}\label{sec:collider}

\begin{figure}
   \begin{center}
		\mbox{
		\subfigure[]{\includegraphics[height=5cm,width=7.55cm]{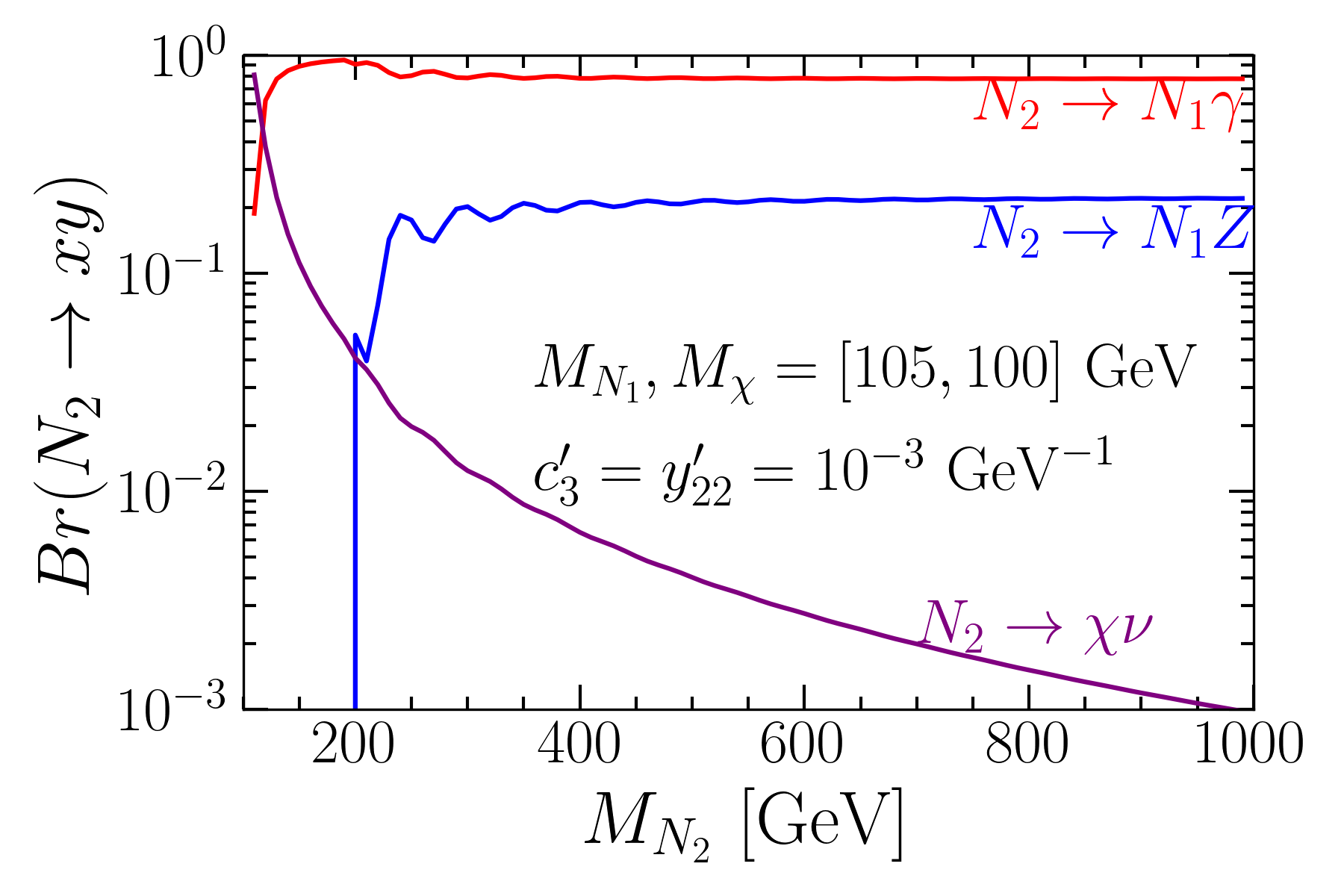}\label{fig:Mvsbr}} 
    \subfigure[]{ \includegraphics[height=5cm,width=7.55cm]{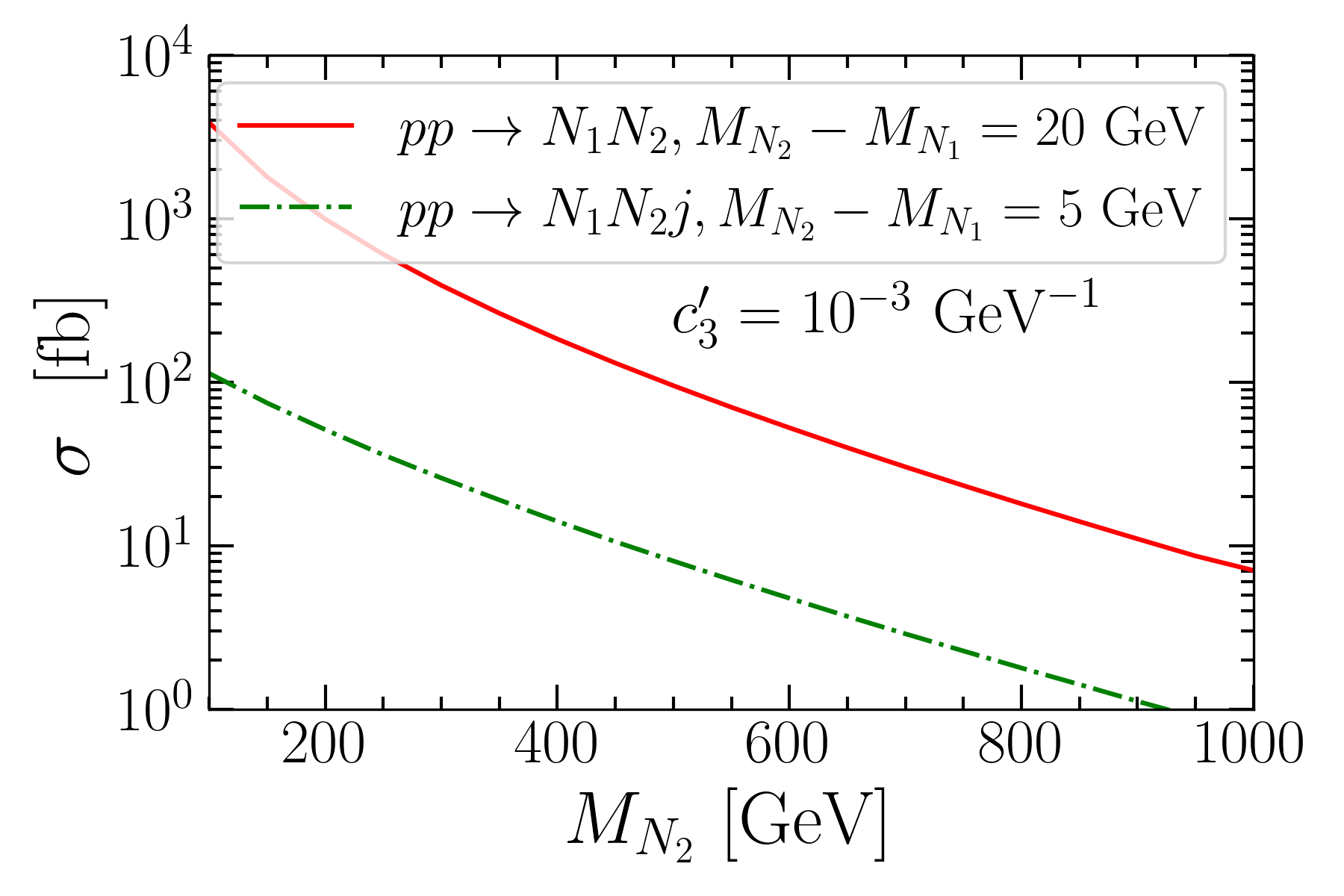}\label{fig:Mvscros}}
 		}
		\caption{(a) Branching ratio of $N_2$ vs its mass. (b) Production cross section of $N_1 N_2$ at 
    $\sqrt{s}=13$ TeV LHC vs $M_{N_2}$. }
		\label{fig:brxs}
	\end{center}
\end{figure}

\begin{figure}
    \begin{center}
		\includegraphics[width=0.45\textwidth]{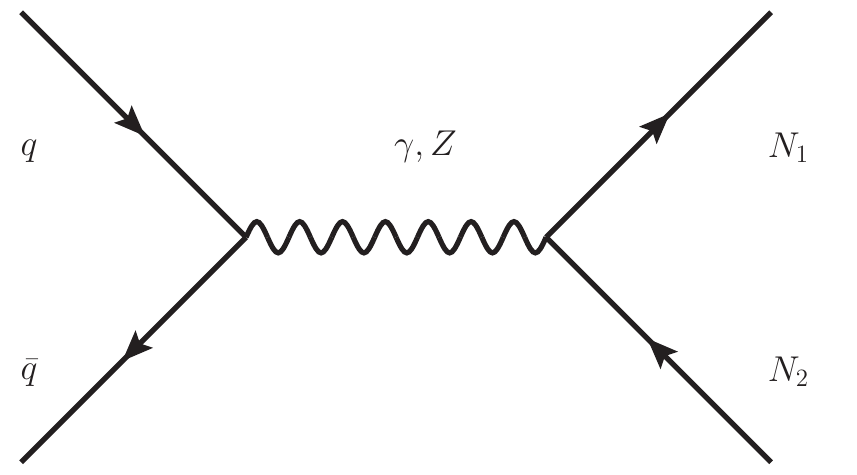}
    \includegraphics[width=0.45\textwidth]{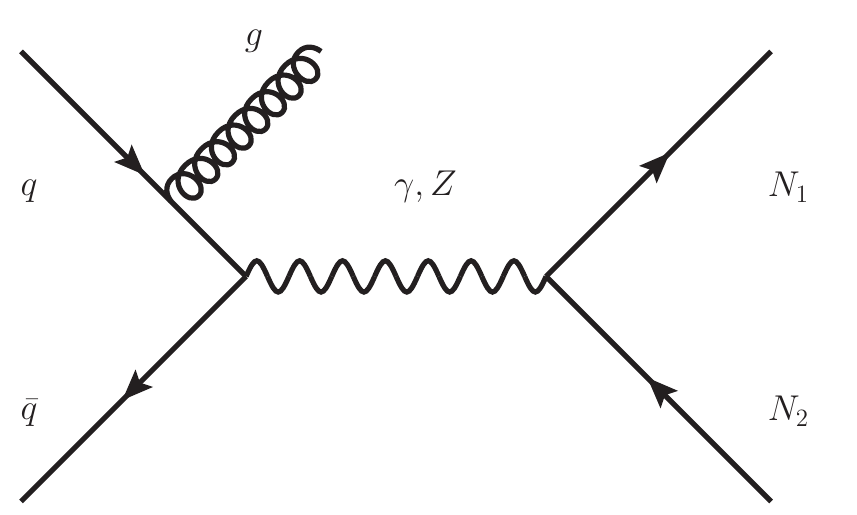}
		\caption{Feynman diagrams for the production of $N_1 N_2$ at pp collider.  }
		\label{fig:feynLHC}
	\end{center}
\end{figure}
In this section we discuss how the LHC can constrain the parameter space of  the model. The singlet fermions  can be produced via the d=5 operators,
 ${c_3^\prime} \overline{N^c} \sigma_{\mu \nu} N B^{\mu \nu}$ with a sizable  rate  at the LHC. The heavier fermion  $N_2$ decays to $N_1 \gamma/N_1 Z$ via the dipole operator or to $\chi \nu$ via Yukawa interactions. 
\begin{figure}
	\begin{center}
	\includegraphics[width=0.6\textwidth]{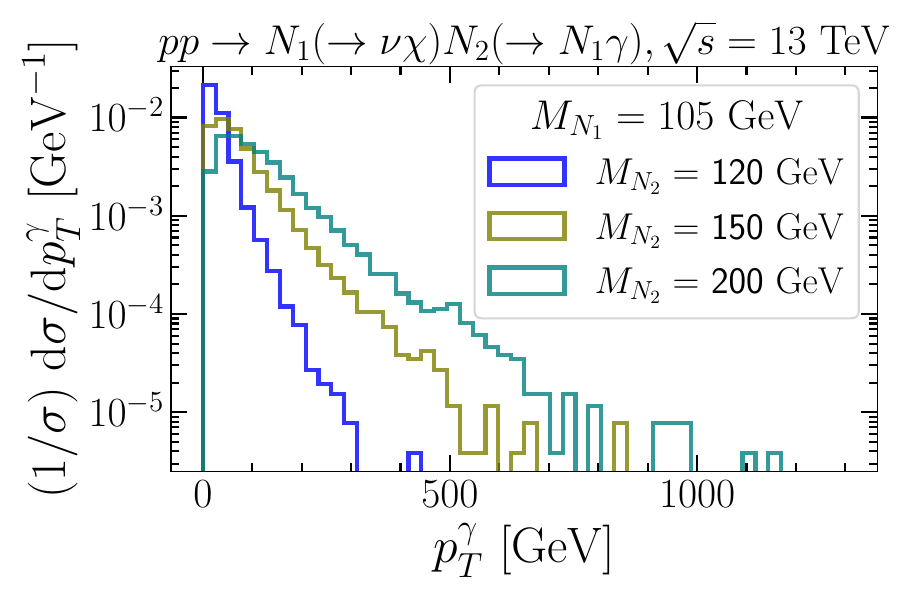}
		\caption{Distribution of the photon $p_T$ for  $M_{N_2}=[120,150,200]$ GeV and $M_{N_1}=105$ GeV.} \label{fig:pT}
	\end{center}    
\end{figure}

Fig.~\ref{fig:Mvsbr} shows the  branching ratio of $N_2$ as function of its mass $M_{N_2}$ for $M_{N_1},M_{\chi}=[105,100]~\rm{GeV}$, $c_{3}^\prime=y_{22}^\prime=10^{-3}~\rm{GeV}^{-1}$. Note that there is no role of the coupling $Y_{11}^\prime$ in the decay of $N_2$.
The decay $N_2 \to N_1 \gamma$ is the dominant channel as soon as it is kinematically accessible while $\text{Br}(N_2\to \chi \nu)\propto 1/{M_{N_2}^2}$ is strongly suppressed as $M_{N_2}$ increases. Moreover $Br(N_2 \to N_1 Z$) is subdominant and quickly saturates to $20\%$. For the choice of the $c_{3}^\prime,y_{22}^\prime$ to be considered (in  Table.~\ref{Table:Scan})for the DM discussion, $\text{Br}(N_2 \to \chi \nu)$ is extremely suppressed. Therefore,  primary decay modes of $N_2$ will be  $N_2\to N_1 \gamma/ N_1 Z$. In Fig.~\ref{fig:Mvscros}, we show the production rate  of $N_1,N_2$ at the LHC. Feynman diagrams for the production processes are shown in Fig.~\ref{fig:feynLHC}.
The dipole operator contributes to $pp\to N_1 N_2$ and for  this dominant production process the cross section scales as $(c_{3}^\prime)^2$.
For $N_1$, the only decay mode is $\chi \nu$. Hence,  the dominant signature of $N_2$ will be mono-$\gamma$ 
associated with large missing momentum while  mono-$Z$ is sub-dominant. The energy of $\gamma/Z$  depend directly on the mass splitting between $N_2$ and $N_1$. When $M_{N_2}-M_{N_1}$ is small, so that the photon is very soft, the mono-jet signal from  $N_1N_2$ associated production with a hard-jet can become important. However, the production cross section for this process drops sharply as we demand high $p_T(j)$.  In Fig.~\ref{fig:Mvscros} we compare the production rate of $pp\to N_1 N_2 j $ (green line) for $p_T(j)\ge200$ GeV with the one for $pp\to N_1 N_2 $ assuming a mass splitting of 5 GeV between  $N_2$ and $N_1$. Note that the mono-jet signal can arise from the process $pp\to N_i N_i j$ via the coupling $c_2^\prime$, here it is irrelevant since 
we set $c_2^\prime=0$. The production of a DM pair via the coupling with Higgs,  $pp\to h \to \chi\chi j$ can also lead to a mono-jet signal but the cross section is suppressed by the coupling $\lambda$  considered, $\lambda <10^{-6}$. We obtain that the mono-jet cross section in our model (for the parameter space of interest) is one order of magnitude smaller than the observed cross section from ATLAS search~\cite{ATLAS:2021kxv}. Therefore mono-$\gamma$ gives the leading signal. We consider only this process  in the following.

\paragraph{Limit from mono -$\gamma$ search:} There exist an ATLAS search for  events with an energetic photon and large missing transverse momentum at the 13 TeV LHC with integrated luminosity of $139~\text{fb}^{-1}$~\cite{ATLAS:2020uiq}. As mentioned above, the production of $N_1 N_2$ via the channel $pp \to N_1 N_2$  mediated by the dipole operator ${c_3^\prime} \overline{N^c} \sigma_{\mu \nu} N B^{\mu \nu}$ can lead to such signal for  the  mass hierarchy $M_{N_2}>M_{N_1}>M_\chi$  such that $N_2 \to N_1 \gamma$ and $N_1$ decay invisibly, $N_1 \to \chi \nu$ leading to $pp\to N_1 N_2 \to \gamma + E_T^{\rm miss}$. The mass splitting $M_{N_2}-M_{N_1}$ is crucial for the sensitivity of the signal as  it governs the energy of the photon. Figure~\ref{fig:pT} shows  the distribution of $p_T^\gamma$ for  $M_{N_2}=[ 120,150,200]$ GeV and $M_{N_1},M_\chi=[105,100]$ GeV. As expected, the photon is harder for larger $M_{N_2}-M_{N_1}$.  The signal regions in the ATLAS search~\cite{ATLAS:2020uiq} demands $E_T^\gamma>150$ GeV and  upper limits at $95\%$ CL on the visible cross section are set for 7 signal regions (SRs): 4 inclusive (SRI1–SRI4) and 3 exclusive (SRE1-SRE3). The SRs are defined corresponding to different  $E_T^{\rm miss} $ ranges,  as listed in Table~\ref{table:SRs}. 

\begin{table}[]
	\centering
	\renewcommand\arraystretch{1.1}
	\begin{tabular}{c| c  c  c  c c c c}
		\hline\hline
		SRs & SRI1  &  SRI2 &  SRI3 &  SRI4 &  SRE1& SRE2 & SRE3 \\
		$E_T^{\rm miss} $ [GeV] & $>200$   & $>250$   & $>300$   & $>375$  & $200-250$ & $250-300$ & $300-375$ \\
		$(\sigma \times A\times\epsilon)_{95}^{\rm obs} $ [fb] &2.45&1.42&0.93&0.53&1.80&1.04&0.79 \\
		\hline\hline
	\end{tabular}
	\caption{Upper limits at $95\%$ CL on the visible cross section $(\sigma \times A\times\epsilon)_{95}^{\rm obs} $  of the signal $pp\to \gamma + E_T^{\rm miss}$ for various signal regions (SRs), where $A$ is the acceptance and $\epsilon$ is the efficiency~\cite{ATLAS:2020uiq}.}
	\label{table:SRs}
\end{table}

\begin{figure}
	\begin{center}
		\mbox{
		\subfigure[]{\includegraphics[width=0.5\textwidth]{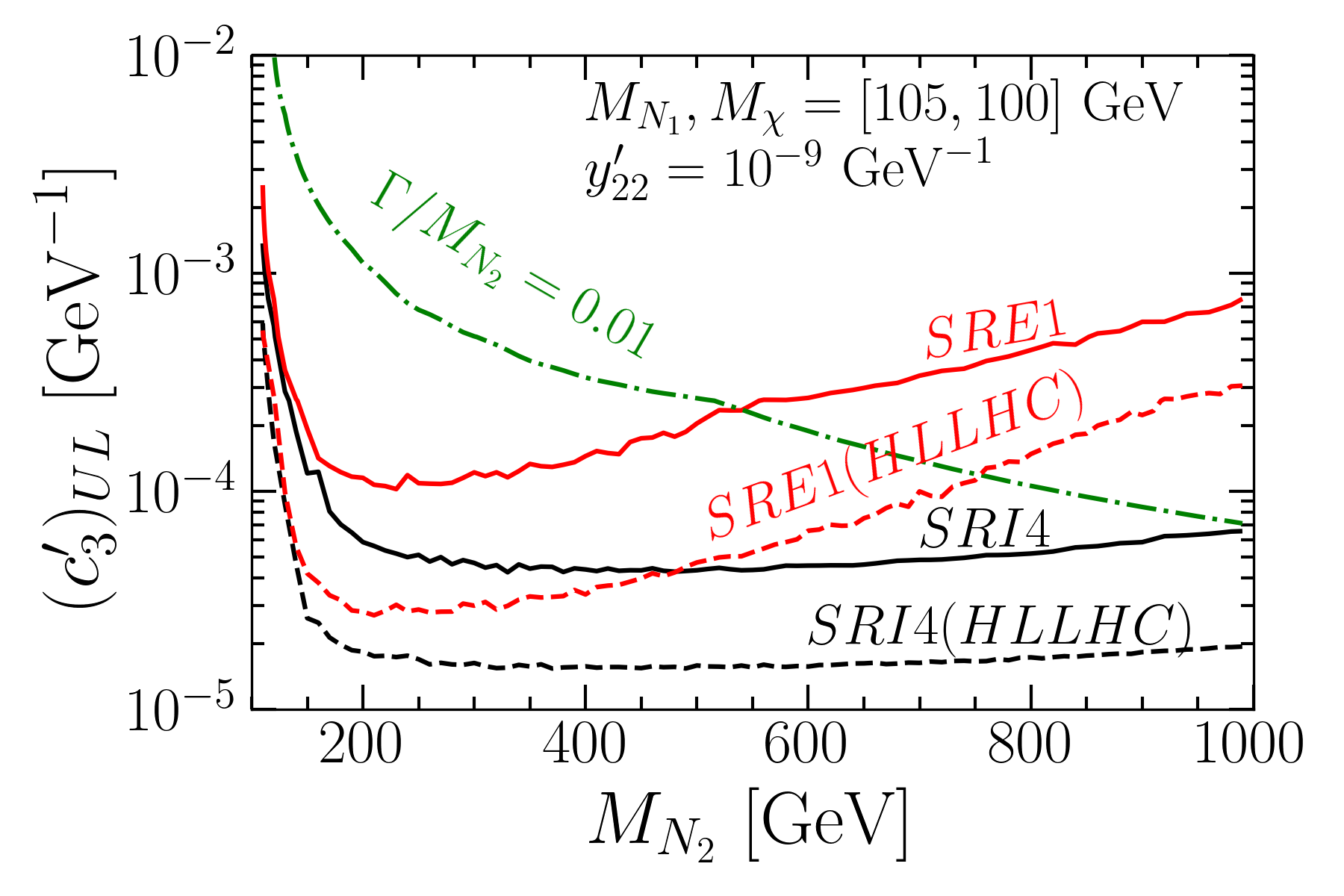}\label{fig:limit}} 
    \subfigure[]{ \includegraphics[width=0.5\textwidth]{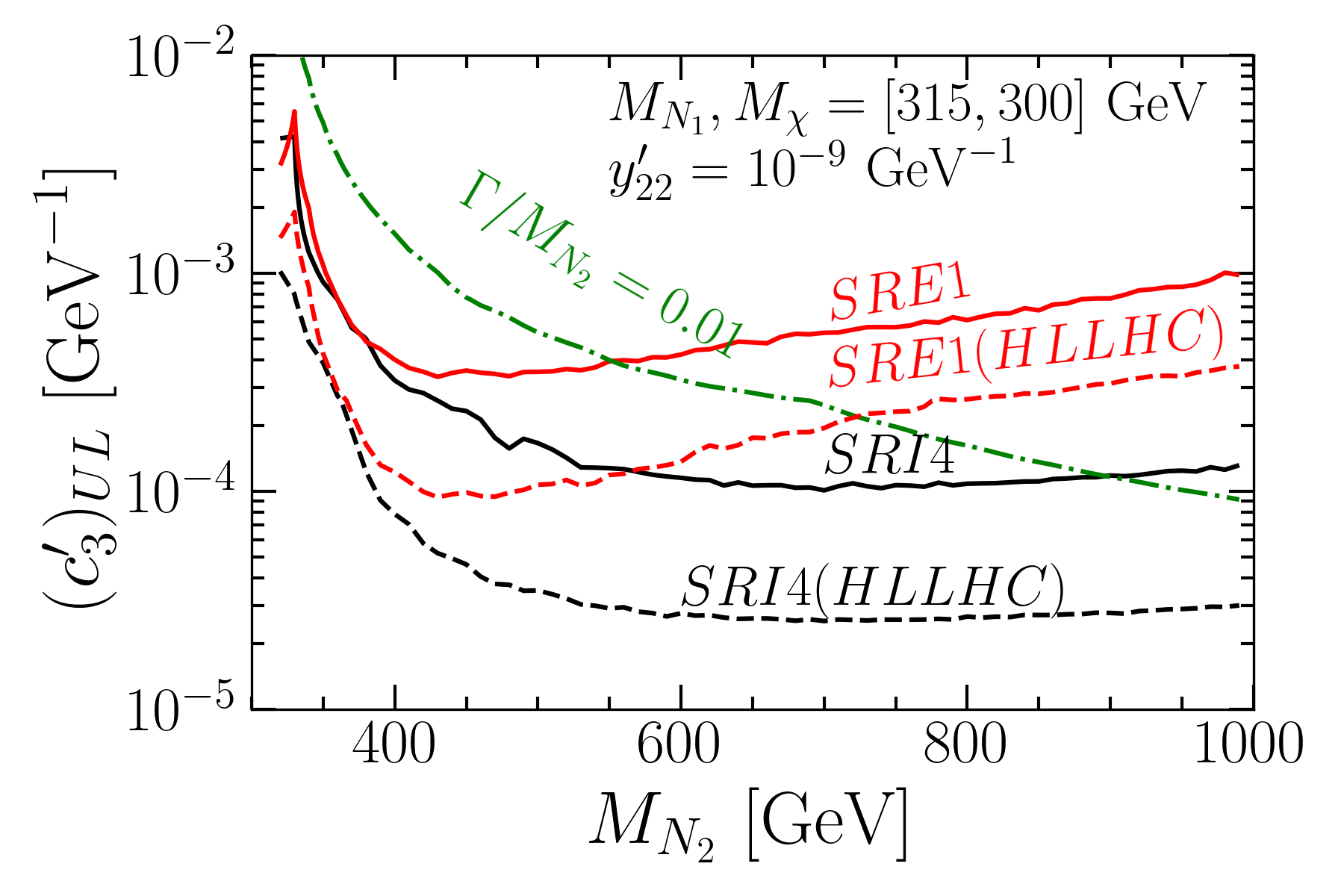}\label{fig:limit300}}
		}

		\caption{Limit from ATLAS search~\cite{ATLAS:2020uiq} in the $(c_3^\prime)_{UL}$ vs $M_{N_2}$ plane for signal region SRE1 (red) and  SRI4 (black) for  (a) $M_\chi=100$ GeV, $M_{N_1}=105$ GeV  (b) $M_\chi=300$ GeV, $M_{N_1}=315$ GeV.  In both cases $y_{22}^\prime = 10^{-9}~{\rm GeV}^{-1}$. }
		\label{fig:lhclimit}
	\end{center}
\end{figure}

\begin{figure}
	\begin{center}
    \mbox{
		\subfigure[]{\includegraphics[width=0.5\textwidth]{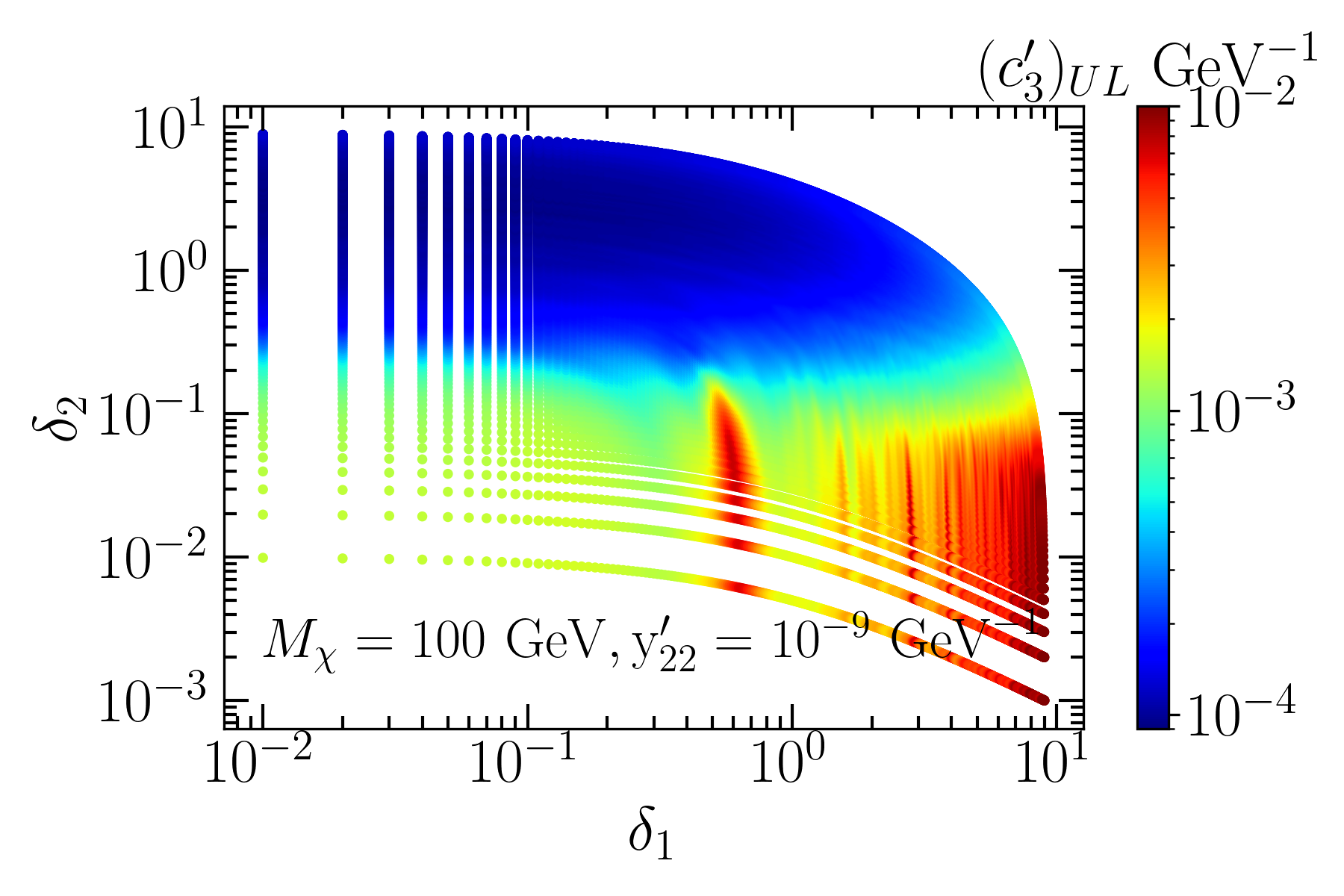}} 
    \subfigure[]{ \includegraphics[width=0.5\textwidth]{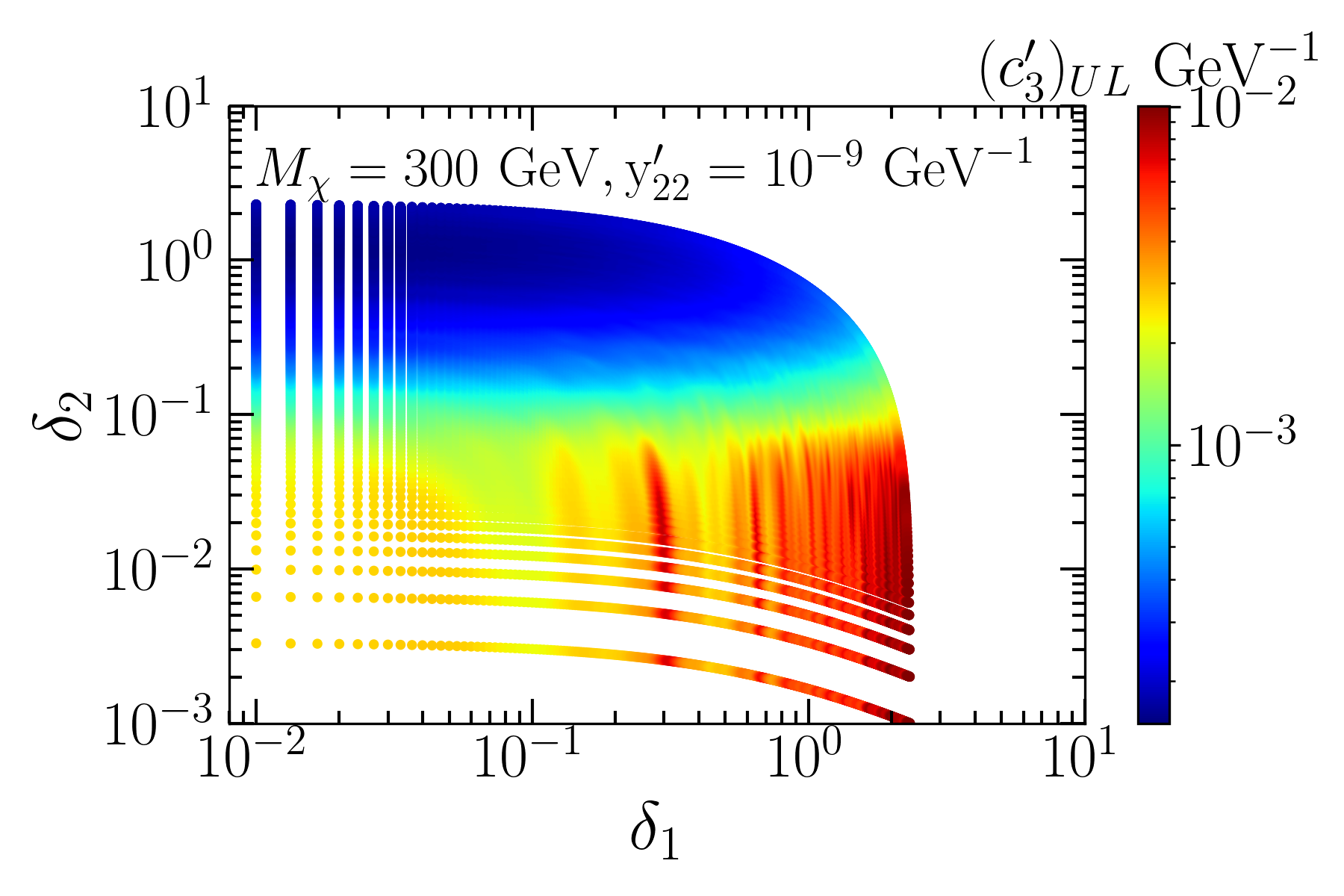}}
		}         
		\caption{Upper limit on $c_3'$ from  the ATLAS search~\cite{ATLAS:2020uiq} in the $\delta_1-\delta_2$ plane for  (a) $M_\chi=100$ GeV and (b) $M_\chi=300$ GeV. Here $y_{22}^\prime = 10^{-9}~{\rm GeV^{-1}}$.}\label{fig:limitdel1del2}
	\end{center}
\end{figure}

The Lagrangian of this model is implemented in FeynRules(v2.3)~\cite{Alloul:2013bka} and the generated UFO files are used in the MC event generator MADGRAPH5(v3.4)~\cite{Alwall:2014hca} to simulate the signal events. Partonic events are passed through PYTHIA8~\cite{Sjostrand:2014zea} to perform showering and hadronization. We use Delphes (3.5)~\cite{deFavereau:2013fsa} for the simulation of detector effect and to calculate the signal efficiencies.

Following the selection cuts used in the ATLAS search~\cite{ATLAS:2020uiq}, we calculated the selection efficiencies for our signal model. We demand that the leading photon has $E_T^\gamma>150$ GeV, $1.52<|\eta^\gamma|<2.37$ or $|\eta^\gamma|<1.37$, and $\Delta \phi(\gamma,E_T^{\rm miss})>0.4$. Then we calculate the signal events in the seven SRs and set limit in the $M_{N_2}$ vs $(c_3^\prime)_{UL}$ plane as shown in Fig.~\ref{fig:lhclimit}, where $(c_3^\prime)_{UL}$ represent the $95\%$ CL upper limit on $c_3^\prime$. 
Here we only show the  strongest and weakest limit. We also calculate the HL-LHC projection, for this we scale the observed upper limit on the visible cross section by a factor $\sqrt{139/3000}$ and compare with the theory prediction.
Fig.~\ref{fig:limit} is for DM mass $M_\chi=100$ GeV, $M_{N_1}=105$ GeV and $y_{22}^\prime=10^{-9}~\rm{GeV}^{-1}$. Here the solid lines indicate current LHC limit while dashed lines are for HL-LHC projection. The strongest limit is set in SRI4 (black)  and the weakest limit in SRE1 (red).  $(c_3^\prime)_{UL}$ is determined by the production cross section $\sigma(pp\to N_1 N_2)$, $Br(N_2 \to N_1 \gamma)$ and cut efficiency $\epsilon$. For a fixed $M_{N_1}$ and $c_3^\prime$, the $\sigma(pp\to N_1 N_2)$ drops with $M_{N_2}$. While, $Br(N_2 \to N_1 \gamma)$ remains steady except near the  kinematic threshold, as evident from Fig.\ref{fig:Mvsbr}. The cut efficiency is different for different SRs. For SRI4, $\epsilon$  increases with $M_{N_2}$ when $M_{N_1}$ is fixed and it can vary from $10^{-4}$ for $M_{N_2}=110$ GeV,  to $0.65$ for $M_{N_2}=1000$ GeV. Whereas, for SRE1, $\epsilon$  increases initially with $M_{N_2}$ and at a certain point it starts dropping.  Therefore, $(c_3^\prime)_{UL}$ varies differently  for different  SRs. The limit  is weakest when $M_{N_2}\simeq M_{N_1}$ as the photon becomes very soft.  For $M_{N_2} \simeq 200$ GeV limit on $c_3^\prime$ becomes strongest. In the region $M_{N_2} > 200$ GeV, for SRI4, the rise in $\epsilon$ gets canceled by the drop in $\sigma$  leading to a  almost flat limit on $c_3^\prime$ w.r.t $M_{N_2}$. Whereas for SRE1, the limit weakens as $\sigma$ drops faster compared to the variation in $\epsilon$.
Fig.~\ref{fig:limit300} shows  the limits  in the case $M_\chi=300$ GeV, $M_{N_1}=315$ GeV and $y_{22}^\prime=10^{-9}~\rm{GeV}^{-1}$ . The  dotted line represents $\Gamma(N_2)/M_{N_2}=0.01$.  Finally, in Fig.~\ref{fig:limitdel1del2}, we show the limit on $c_3^\prime$ in the $\delta_1$ vs $\delta_2$ plane, as defined  in Eq.~\ref{eq:relative_mass_splitting}. 
Here we consider only SRI4 since it  sets the strongest limit. 
\begin{figure}
	\begin{center}
		\mbox{
			\subfigure[]{\includegraphics[height=6.5cm,width=8.00cm]{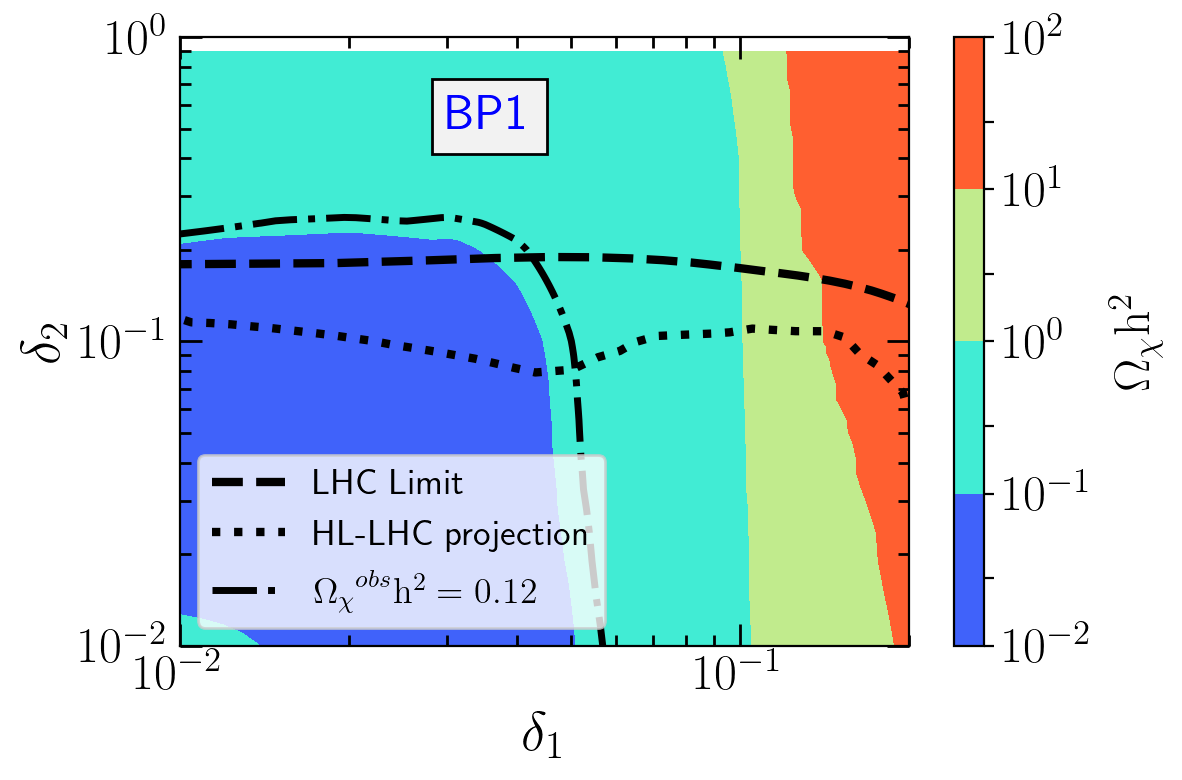}\label{fig:sc_1a_scan_delta_res1}} 
			\subfigure[]{\includegraphics[height=6.5cm,width=8.00cm]{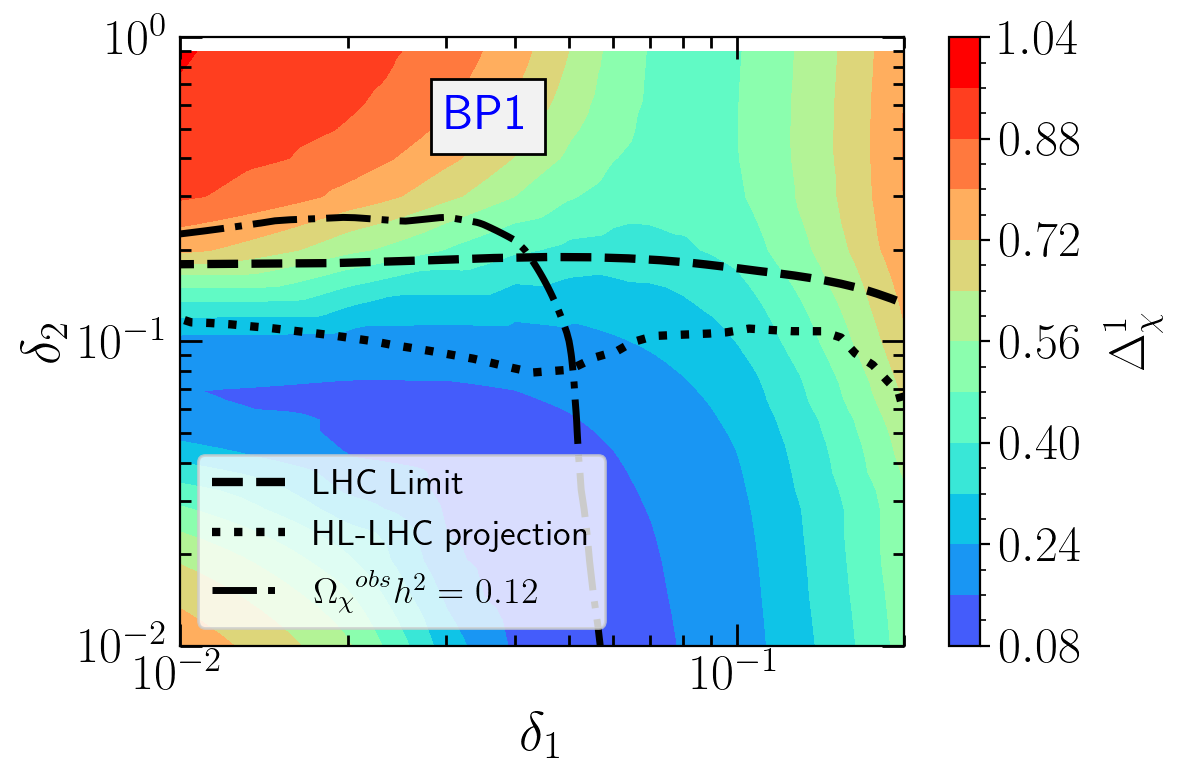}\label{fig:sc_1a_scan_delta_res2}}		}
		\caption{(a) DM relic density and  (b) $\Delta^1_\chi$ in the  $\delta_1$ and $\delta_2$ plane for $M_{\chi}=100\,\text{GeV}$, $c_{3}^{\prime}=7\times 10^{-4}\,\rm{GeV^{-1}}, y^{\prime}_{11,22}=1.1\times10^{-9}\,\rm{GeV^{-1}}$. The dash- dotted line satisfies the DM observed relic density.
        }
		\label{fig:sc_1a_scan_delta_res}
	\end{center}
\end{figure}

\section{Numerical Analysis}\label{sec:Numerical_Analysis}
As discussed  in Section~\ref{sec:FO}, we re-emphasize that the DM $\chi$ gets diluted in the early Universe either through co-annihilation, which is primarily governed by $N_{1,2}$ decoupling from the thermal bath or through the inelastic processes co-scattering. The inverse decay process primarily governs the inelastic processes, i.e. $\chi \nu_{e,\mu} \to N_{1,2}$. To delineate the parameter space of our model, we first concentrate on the two benchmark points, \textbf{BP1} and \textbf{BP2}, introduced in Section~\ref{sec:FO} but with varying $M_{N_{1}}$ and $M_{N_{2}}$ and keeping other parameters fixed. This is followed by a comprehensive exploration of the model’s parameter space.

In Fig.~\ref{fig:sc_1a_scan_delta_res1}, we show the variations of the DM relic density contours in the $\delta_{1}-\delta_{2}$ plane for BP1. The black dashed-dot line corresponds to $\Omega_{\chi}h^{2}=0.12$.   When $\delta_{2}$ is less than $0.15$ the contour line corresponding to $\Omega_{\chi}h^{2}=0.12$ is approximately independent of $\delta_{2}$. It arises because of efficient dilution of $N_{1,2}$ through co-annihilation as $\delta_{2}$ decreases, and the relic density is set  when $N_{1}$ decouples through the thermal bath.  Figure~\ref{fig:sc_1a_scan_delta_res2} shows that $\Delta^{1}_{\chi}$ decreases as $\delta_2$ decreases. Hence, in this limit, the standard freeze-out via co-annihilation is realized. On the contrary, for $\delta_{1}$ less than $5\times10^{-2}$, the inverse decay rate is suppressed due to phase space suppression and satisfies the condition given in Eq.~\ref{eqn:inv_decay_condition}. Furthermore, the dilution of $N_{1,2}$ is not efficient owing to the large mass splitting between $N_{1}$ and $N_2$. Hence, the relic density gets frozen when the inverse decay of $\chi$ to $N_{1}$ ceases. It is in the spirit of a co-scattering mechanism, and $\Delta^{1}_{\chi}$ approaches 1. Also, note that DM is overabundant  for large $\delta_1$ and $\delta_2$  due to the fact that neither the inverse decay nor (co-)annihilation of $N_{1,2}$ are efficient. The black dashed and dotted line show the current LHC limit and HL-LHC future projection in the $\delta_{1}-\delta_{2}$ plane. As illustrated in Fig.~\ref{fig:sc_1a_scan_delta_res2}, the co-scattering dominated region is subject to stringent constraints from LHC~\cite{ATLAS:2021kxv}, with only values of $\Delta^{1}_{\chi}\lesssim0.4$ allowed when imposing that $\Omega h^2=0.12$. The HL-LHC has the potential to probe the co-scattering region when $\Delta^{1}_{\chi} > 0.2$ .

\begin{figure}
	\begin{center}
		\mbox{
			\subfigure[]{\includegraphics[height=6.5cm,width=8.00cm]{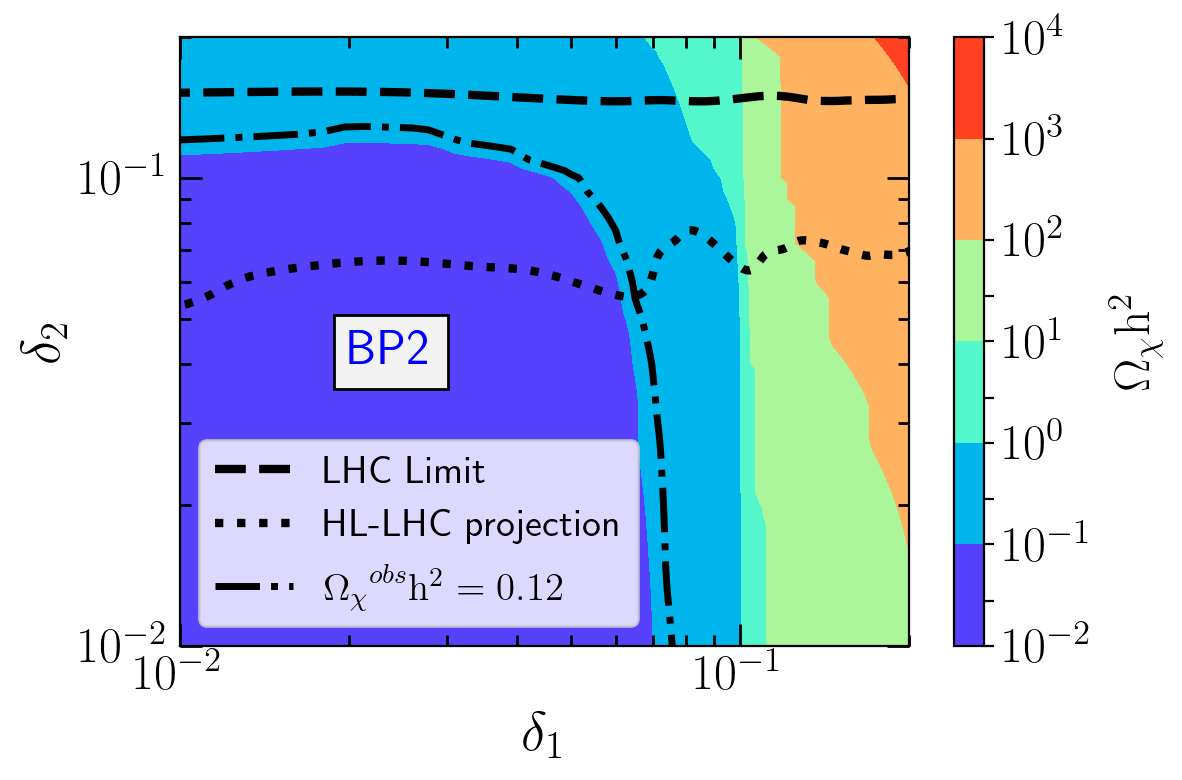}\label{fig:sc_300_scan_delta_res1}} 
			\subfigure[]{\includegraphics[height=6.5cm,width=8.00cm]{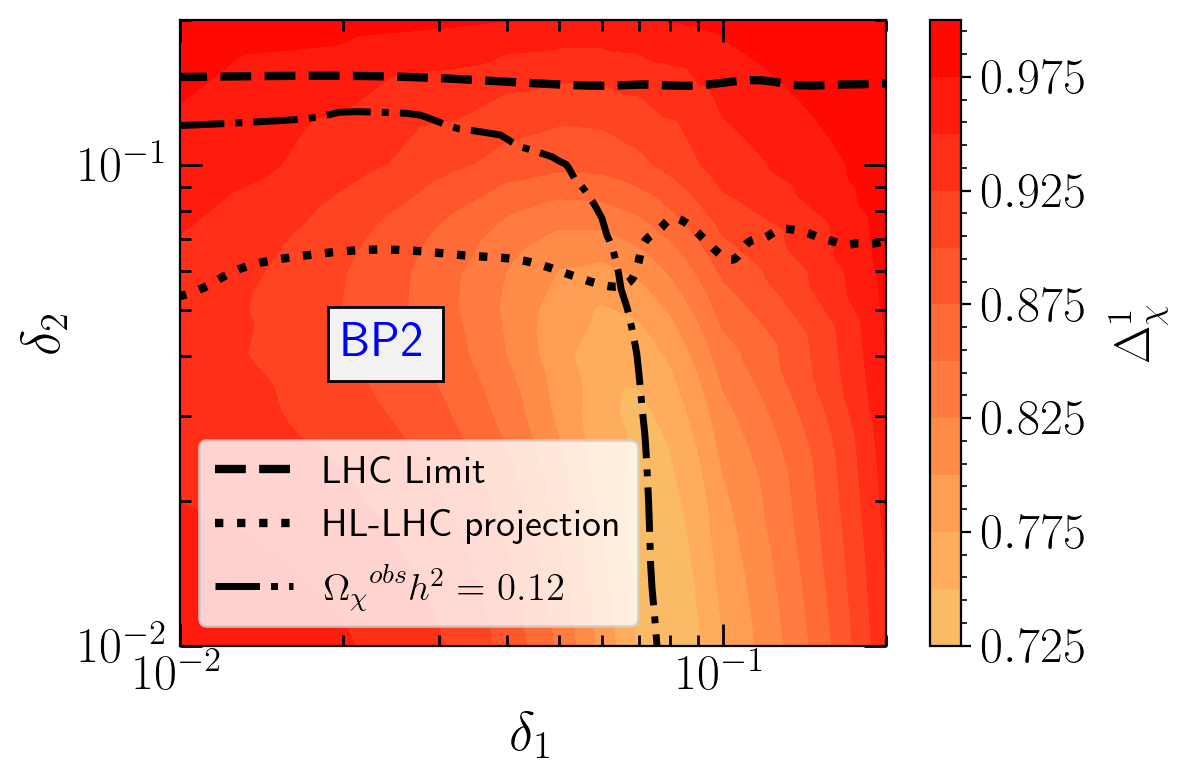}\label{fig:sc_300_scan_delta_res2}}		}
		\caption{ Same as Fig.~\ref{fig:sc_1a_scan_delta_res} for $M_{\chi}=300\,\text{GeV}$. }
		\label{fig:sc_1a_scan_delta_res_300}
	\end{center}
\end{figure}

\begin{table}[]
	\centering
	\begin{tabular}{| c | c  | }
		\hline
		{Parameter}& \multicolumn{0}{ c |   }{\quad Scanned range }   \\
		\hline
		$M_{\chi}\,\,\rm{[GeV]}$ & \quad [$100$ , $500$]         \\                  
		$\delta_{1} (M_{N_1}~\rm{[GeV]})$       &   \quad [$10^{-3}$ , $0.5$]  $([\approx M_\chi, 750])$       \\
        $\delta_{2} (M_{N_2}~\rm{[GeV]})$       &   \quad [$10^{-3}$ , $1$] $([\approx M_{N_1},1500])$         \\
		$y^{\prime}_{11}\,\,\rm{[GeV^{-1}]}$       &   \quad [$10^{-11}$ , $10^{-6}$]                \\
		$y^{\prime}_{22}\,\,\rm{[GeV^{-1}]}$       &   \quad [$10^{-11}$ , $10^{-6}$]   \\
		$c^{\prime}_{3}\,\,\rm{[GeV^{-1}]}$       &   \quad [$10^{-6}$ , $10^{-3}$]                   \\
		\hline
	\end{tabular}
	\caption{\centering Input parameters used in the numerical scan to determine  the allowed parameter space. Note that we have assumed $c^{\prime}_{1,2}=0$ and $\lambda=10^{-6}$.} 
	\label{Table:Scan}
\end{table}
\begin{figure}
	\begin{center}
		\mbox{
		\subfigure[]{\includegraphics[width=0.52\textwidth]{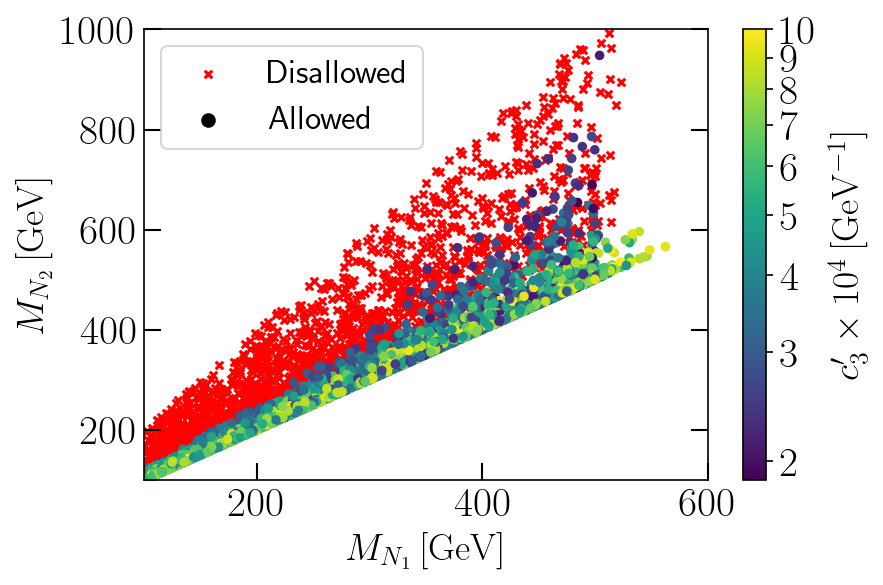}\label{fig:Collider_Allowed_Scan}} 
    \subfigure[]{ \includegraphics[width=0.52\textwidth]{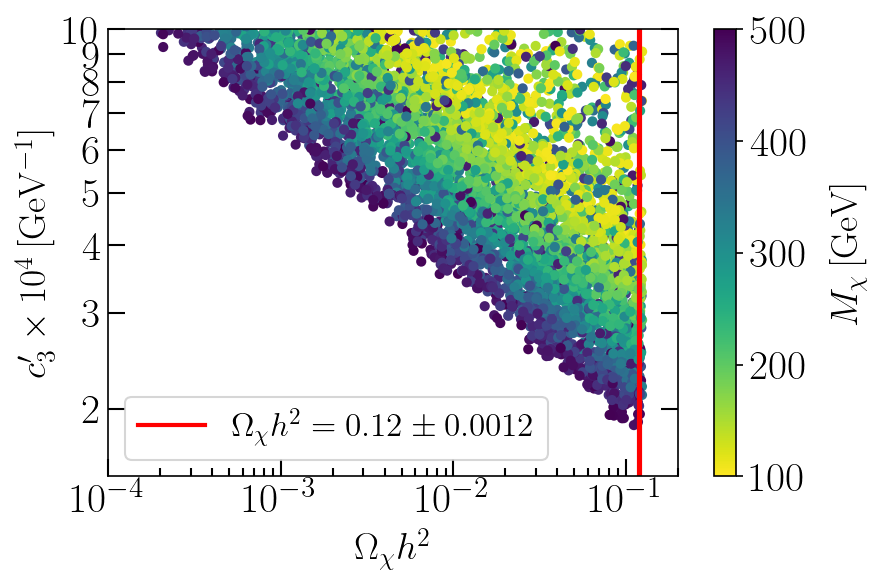}\label{fig:Relic_c312_plane_Scan}}
		}
	\mbox{\subfigure[]{\includegraphics[width=0.52\textwidth]{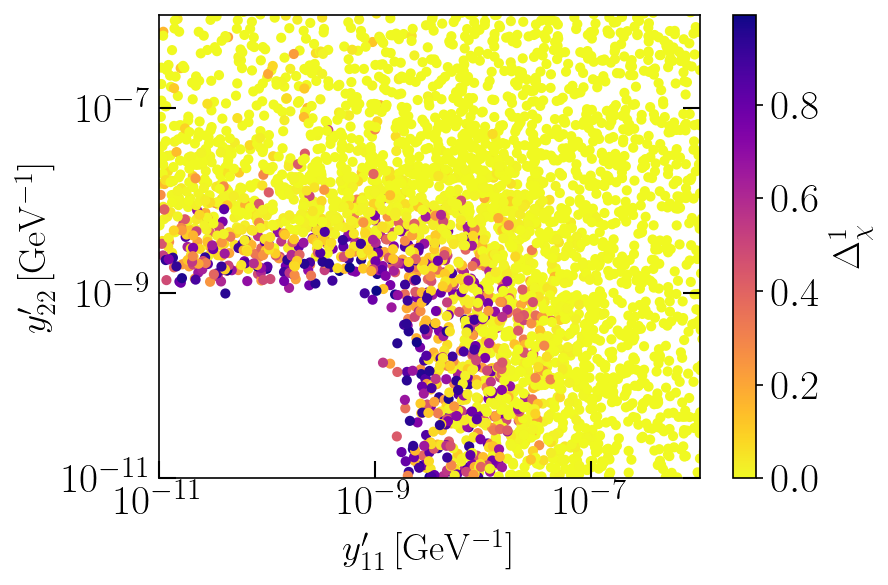}\label{fig:Y11_Y22_plane_Delta_Scan} }
     \subfigure[]{ \includegraphics[width=0.52\textwidth]{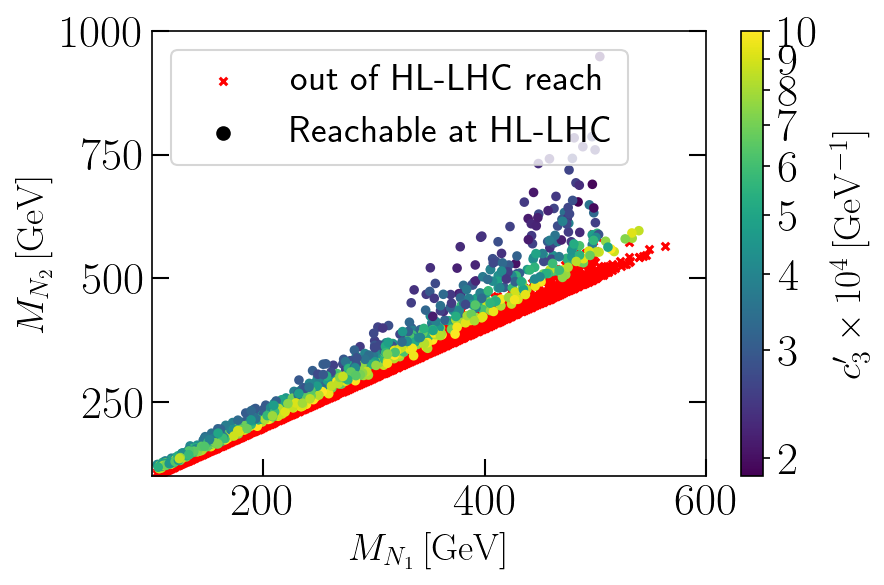}\label{fig:future_collider}}
    }
	\caption{Fig.~\ref{fig:Collider_Allowed_Scan} illustrates (dis)allowed parameter space in $M_{N_{1}}-M_{N_{2}}$ plane, subject to relic density constraints $10^{-4}\leq\Omega_{\chi}h^{2}\leq0.1224$ along with collider limit. The color pallet corresponds to $c^{\prime}_{3}$. Fig.~\ref{fig:Relic_c312_plane_Scan} shows variations of $\Omega_{\chi}h^{2}$ vs $c^{\prime}_{3}$ with the DM mass $M_{\chi}$ in color pallet after imposing the LHC constraint. Fig.~\ref{fig:Y11_Y22_plane_Delta_Scan} shows variations of $y^{\prime}_{11}$ vs $y^{\prime}_{22}$ with the $\Delta^{1}_{\chi}$ in the color pallet. Note that we have assumed $c^{\prime}_{1,2}=0$ and $\lambda=10^{-6}$. Fig.~\ref{fig:future_collider} Shows parameter space that can be probed at the HL-LHC.
    }
		\label{fig:Gobal_scan_S1}
	\end{center}
\end{figure}

Following the same approach as above, we show in Fig.~\ref{fig:sc_1a_scan_delta_res_300}   the contours of DM $\chi$ relic density and $\Delta^{1}_{\chi}$ for $M_{\chi}=300\,\rm{GeV}$. As before, we have fixed $c^{\prime }_{3}=7\times10^{-4}\,\rm{GeV^{-1}}$. However, the allowed masses for $N_1$  and $N_2$  in \textbf{BP2} are larger than those in \textbf{BP1} to ensure that $\chi$ be the DM candidate. This results in a decrease in $N_{1,2}$ (co-)annihilation cross-section for \textbf{BP2} compared to \textbf{BP1} and, in turn, facilitates early $N_{1,2}$ decoupling from the thermal bath in \textbf{BP2}. Therefore, the relic density of $\chi$ is primarily governed by the inverse decay process of $\chi$ to $N_{1,2}$ and results in a large value of $\Delta^{1}_{\chi}$, see Fig.~\ref{fig:sc_300_scan_delta_res2}. In this case, the LHC does not constrain the parameter space where the relic density  matches the observed value as it requires a too small value of $\delta_2$, as seen from the black dashed and dot-dashed  lines in Fig.~\ref{fig:sc_300_scan_delta_res2}. However the  HL-LHC can probe some of the parameter space in the co-scattering dominant region. 

We comprehensively explored the model’s parameter space through a flat random scan over the free parameters within the ranges shown in Table~\ref{Table:Scan}. We consider the points which satisfy the DM relic density in the range,
\begin{eqnarray}
    10^{-4}\leq\Omega_{\chi}h^{2}\leq0.1224.
    \label{eqn:relic_range_scan}
\end{eqnarray}

In Fig.~\ref{fig:Collider_Allowed_Scan}, we show the allowed range  of $M_{N_{1}}$ vs $M_{N_{2}}$  with $c^{\prime}_{3}$ in the color pallet after imposing the relic density constraint as given by Eq.~\ref{eqn:relic_range_scan}. We then impose 
the ATLAS mono-$\gamma$ limit as discussed in Sec.~\ref{sec:collider}.  Figure ~\ref{fig:Collider_Allowed_Scan} shows that the 
mono-$\gamma$ search strongly constrain scenarios with 
 large mass splitting between $M_{N_1}$ and $M_{N_2}$, these points are shown in red in the figure. Furthermore, as discussed above,  the relic density constraint requires at least $N_1$ to be present in the early Universe, so that $\chi$ can be diluted through co-annihilation or inverse decay. Thus the mass difference between  $M_{N_1}$ and $M_{\chi}$ cannot be too large, and we obtain an upper limit on $M_{N_1}<600\,\rm{GeV}$.  It is important to notice  that $N_{2}$ can also participate in DM $\chi$ dilution through co-annihilation or inverse decay. However since the contribution of $N_1$ is always present, the upper bound  on $M_{N_2}$ is less stringent, specifically, $M_{N_2} \leq 1000\,\mathrm{GeV}$.
 
In Fig.~\ref{fig:Relic_c312_plane_Scan}, we depict the allowed points after  the relic density constraint and collider limit in the $\Omega_{\chi}h^{2}$ and $c^{\prime}_{3}$ plane with the color pallet corresponding to the DM mass $M_{\chi}$. The points overlapping with the red line satisfy the observed DM relic density within the $2\sigma$ range. As $c^{\prime}_{3}$ decreases, $N_{1,2}$ undergoes early chemical decoupling from the thermal bath, which in turn does not contribute in $\chi$ dilution neither through co-annihilation nor through co-scattering. We can see that for the range of $M_{\chi}$ that survived $c^{\prime}_{3}\leq 3\times10^{-4}$ , i.e. $M_{\chi}>200\,\rm{GeV}$, corresponds to $M_{N_{1,2}}>M_{\chi}$. It happens because, for $M_{N_{1,2}}>200\,\rm{GeV}$ additional channels  such as  $N_{1}N_{2}\to t\bar{t}$ become kinematically accessible. In Fig.~\ref{fig:Y11_Y22_plane_Delta_Scan}, we show the variation of the allowed points in $y^{\prime}_{11}$ - $y^{\prime}_{22}$ plane with the color pallet corresponding to $\Delta^{\chi}_{1}$. As evident from Fig.~\ref{fig:Y11_Y22_plane_Delta_Scan}, for large $y^{\prime}_{11,22}$, $\chi $ thermalizes with the thermal bath due to rapid decay and inverse decay as  $\Delta^{\chi}_{1}\to 0$, and the relic density is governed by the thermal decoupling of $N_{1,2}$. It is the region of parameter space where a co-annihilation process governs $\Omega_{\chi}h^2$. On the  contrary, for $y^{\prime}_{11,22}<10^{-8}\,\rm{GeV^{-1}}$, $\chi$  undergoes early chemical decoupling and $\Omega_{\chi}h^2$ gets frozen when $\chi$ inverse decay to $N_{1,2}$ ceases. 
Consequently, in this region of parameter space, the value of $\Delta^{\chi}_{1}>0.5$, indicating that the DM relic density is predominantly determined by the co-scattering mechanism.
In Fig~\ref{fig:future_collider}, we show the discovery prospects via the mono-$\gamma$ signal at the HL-LHC. The dotted-color points in the $M_{N_{1}}$ and $M_{N_{2}}$ plane can be probed in HL-LHC while the red-crossed points remain out of HL-LHC reach. These points correspond to lower values of  $M_{N_{2}}-M_{N_{1}}$ and/or lower value of $c^{\prime}_{3}$, which leads to suppressed signal efficiency and/or suppressed production cross section for the signal $pp\to N_1 N_2 \to \gamma +E_T^{\rm miss}$.

\section{Summary and Conclusion}\label{sec:conclusion}
In this work, we investigate the dark matter production mechanism in an  EFT framework where we extent the SM with two singlet fermions $N_{1,2}$ and a real singlet scalar $\chi$, which are $Z_2$ odd particles. The singlet scalar $\chi$ serves as a dark matter candidate.  Both $\chi$ and $N_{1,2}$ are thermal particles  and their interactions in the early Universe controls the DM abundance. We emphasize the role of dimension-5 operators in the co-scattering mechanism.  The key parameters on which DM dynamics depend are the masses $M_{N_{1,2}},M_\chi$ and couplings $\lambda,c_3^\prime, y^\prime_{11,22}$. Note that, we have assumed $c_{1,2}^\prime=0$ throughout our analysis.

We first discuss DM formation using two benchmarks, BP1 where the co-annihilation mechanism dominates and BP2 where co-scattering dominates.  We fix the Higgs portal coupling $\lambda=10^{-6}$ ensuring that  $\chi$ remains in thermal equilibrium and also to evade stringent direct detection constraints.
 The DM relic density  in {\bf BP1} is set when $N_1$ decouples from the thermal bath and  the abundance is determined by the freeze-out of processes that maintain equilibrium between states until decoupling. In contrast, for {\bf BP2}, the relic density is set when the inverse decay process ceases and is governed by the co-scattering mechanism, where conversions between states—rather than annihilations—control the final abundance. In both cases we find that the inverse decay rates,  $\chi, \nu_{e,\mu} \to N_{1,2}$, although more efficient in {\bf BP1}, are not sufficient to bring $\chi$ into chemical equilibrium with the thermal bath. We also find as expected  that small mass splittings ($\delta_1<0.1$ or $\delta_2<0.1$) are needed to prevent overproduction of DM. 
 
 As concerns the collider constraints,  we find that  the dimension-5 operators lead to sizable production of $N_{1,2}$ at the LHC and that the most stringent constraint is coming from the mono-$\gamma$ signature via $pp \to N_1 N_2 (\to N_1 \gamma) \to \gamma +E_T^{\rm miss}$. Following the ATLAS analysis, we calculate the sensitivity of this signal  which  primarily depends on the mass difference $M_{N_2}-M_{N_1}$ (or the photon energy) and $c_3^\prime$. We calculate the mono-$\gamma$ limit on  $c_3^\prime$ in the $\delta_2-\delta_1$ plane, where $\delta_1$ dictates the  relative mass splitting between $M_{N_1}$ and $M_\chi$. The stronger bound  on $c_3^\prime$  is found for larger  $\delta_2$ since the energy of the photon grows with $\delta_2$.

We also performed a comprehensive analysis of the model parameter space to identify the regions that satisfy both the relic density and the collider constraints. We found a significant number of points satisfying the observed dark matter relic density and collider limit within the mass range of $100$–$500\,\mathrm{GeV}$. For larger couplings, $y^\prime_{11,22} \ge 10^{-6}~\text{GeV}^{-1}$, the DM $\chi$ and $N_{1,2}$ remain in thermal equilibrium, which corresponds to the usual freeze-out mechanism. However, for smaller couplings $y^\prime_{11,22} \le 10^{-6}~\text{GeV}^{-1}$, chemical decoupling of $\chi$ might occur at early epoch and its abundance is set by the co-scattering process. We determine the potential of HL-LHC to probe further the regions where co-scattering dominates.

In conclusion, we have investigated the production of dark matter, with a primary focus on the co-scattering mechanism within the framework of EFT.  This framework offers rich dark matter phenomenology that can be comprehensively probed at the HL-LHC. While we concentrate on the dipole operator, other   d=5 operators  could play a crucial role for dark matter production, these will be studied in future work.

\section*{Acknowledgements}
The work of AR is supported by Basic Science Research Program through the National Research Foundation of Korea(NRF) funded by the Ministry of Education through the Center for Quantum Spacetime (CQUeST) of Sogang University (RS-2020-NR049598). RP is supported in part by Basic Science Research Program through the National Research Foundation of Korea (NRF) funded by the Ministry of Education, Science and Technology (NRF-2022R1A2C2003567).
GB thanks IFIRSE, Quy Nhon, Vietnam and {\it Rencontres du Vietnam} for their warm hospitality while this work was finalised. This work was funded in part by the Indo-French Centre for the Promotion of Advanced Research (Project title: Beyond Standard Model Physics with Neutrino and Dark Matter at Energy, Intensity and Cosmic Frontiers, Grant no: 6304-2). The authors also acknowledges SAMKHYA: High-Performance Computing Facility provided by the Institute of Physics (IoP), Bhubaneswar.
\bibliographystyle{JHEP}
\bibliography{bibitem.bib}
\end{document}